**Title:** Analysis of cosmic rays' atmospheric effects and their relationships to cutoff rigidity and zenith angle using Global Muon Detector Network data

**Authors:** R. R. S. Mendonça[1,2]; C. Wang[1]; C. R. Braga[2]; E. Echer[2]; A. Dal Lago[2]; J. E. R. Costa[2]; K. Munakata[3]; H. Li[1]; Z. Liu[1]; J.-P. Raulin[4]; T. Kuwabara[5]; M. Kozai[6]; C. Kato[3]; M. Rockenbach[2]; N. J. Schuch[2]; H. K. Al Jassar[7]; M. M. Sharma[7]; M. Tokumaru[8]; M. L. Duldig[9]; J. E. Humble[9]; P. Evenson[10]; and I. Sabbah[11].

**Authors Afiliations:** [1] State Key Laboratory of Space Weather, National Space Science Center (NSSC), Chinese Academy of Sciences, NO. 1 Nanertiao, Zhongguancun, Beijing, 100190, People's Republic of China; [2] National Institute for Space Research, São José dos Campos, SP, 12227-010, Brazil; [3] Physics Department, Shinshu University, Matsumoto, Nagano, 390-8621, Japan; [4] Center of Radio Astronomy and Astrophysics Mackenzie, Engineering School, Mackenzie Presbyterian University, São Paulo, Brazil; [5] Graduate School of Science, Chiba University, Chiba City, Chiba 263-8522, Japan; [6] Institute of Space and Astronautical Science, Japan Aerospace Exploration Agency (ISAS/JAXA), Sagamihara, Kanagawa 252-5210, Japan; [7] Physics Department, Kuwait University, Kuwait City, 13060, Kuwait; [8] Solar Terrestrial Environment Laboratory, Nagoya University, Nagoya, Aichi, 464-8601, Japan; [9] School of Natural Sciences, University of Tasmania, Hobart, Tasmania, 7001, Australia; [10] Bartol Research Institute, Department of Physics and Astronomy, University of Delaware, Newark, DE 19716, USA; [11] Department of Applied Sciences, College of Technological Studies, Public Authority for Applied Education and Training, Shuwaikh, Kuwait.

**Key points**:

- Experimental analysis of pressure and temperature effects observed by ground muon detectors in relation to cutoff rigidity and zenith angle;
- The best correlation was found when considering product between cutoff rigidity and zenith angle secant (pressure) or cosine (temperature);
- The temperature effect only shows a global trend if a relationship with the sine of each detector's geographic latitude is included.

**Abstract:** Cosmic rays are charged particles whose flux observed at Earth shows temporal variations related to space weather phenomena and may be an important tool to study them. The cosmic ray intensity recorded with ground-based detectors also shows temporal variations arising from atmospheric variations. In the case of muon detectors, the main atmospheric effects are related to pressure and temperature changes. In this work, we analyze both effects using data recorded by the Global Muon Detector Network (GMDN), consisting of four multidirectional muon detectors at different locations, in the period between 2007 and 2016. For each GMDN directional channel, we obtain coefficients that describe the pressure and temperature effects. We then analyze how these coefficients can be related to the geomagnetic cutoff rigidity and zenith angle associated with cosmic-ray particles observed by each channel. In the pressure effect analysis, we found that the observed barometric coefficients show a very clear logarithmic correlation with the cutoff rigidity divided by the zenith angle cosine. On the other hand, the temperature coefficients show a good logarithmic correlation with the product of the cutoff and zenith angle cosine after adding a term proportional to the sine of geographical latitude of the observation site. This additional term implies that the temperature effect measured in the northern hemisphere detectors is stronger than that observed in the southern hemisphere. The physical origin of this term and of the good correlations found in this analysis should be studied in detail in future works.

## 1 - Introduction

Cosmic rays are charged particles (mostly protons) with energies from MeV to ZeV ($10^{21}$ eV) that hit Earth's atmosphere almost isotropically. Particles with energies up to a few tens of GeV move in the interplanetary medium responding to the dynamic and magnetic variations of the solar wind plasma (Moraal, 2013). In this way, when observing these particles, we can notice temporal variations of cosmic ray intensity related to solar and interplanetary phenomena (Bazilevskaya, 2000; Kudela, 2009). Many studies have been done about long-term variations related to the 11 and 22 year solar cycles and short-term variations related to solar/interplanetary phenomena as Solar Energetic Particles Events, Coronal Mass Ejections and High-Speed Solar Wind Streams (Cane, 2000; Ryan et al., 2000; Singh and Badruddin, 2007; Potgieter, 2013). Beyond increasing knowledge about high-energy particles and space plasma physics, analysis of relations between cosmic ray intensity variations and solar/interplanetary phenomena can help developing space weather forecast and monitoring tools (Munakata et al., 2000; Belov at al., 2003; Leerungnavarat et al., 2003; Kudela and Storini, 2006; Rockenbach et al. 2014; Dorman, 2012; Papailiou et al., 2012).

Besides the extra-terrestrial influences, Earth's magnetic field and atmosphere can also affect the cosmic ray intensity observed at ground level. When primary cosmic rays in space approach Earth, they interact with the geomagnetic field. Depending on their rigidity, their trajectories are more or less deflected by this field. In this way, knowing the geomagnetic field configuration, we can calculate the geomagnetic cutoff rigidity, which corresponds to the minimum rigidity of primary particles that can arrive at a given location on the Earth's surface and from a given direction (Smart et al. 2000, Herbst et al. 2013). After the interaction with the geomagnetic field, primary cosmic rays that continue moving towards Earth's surface will interact with atmospheric nuclei generating secondary particles including muons and neutrons (Grieder, 2001). Therefore, when measuring the cosmic ray intensity at ground, we observe temporal variations related to time changes in some atmospheric parameters (Dorman, 2004). The way in which this occurs depends on the kind of secondary particle we are observing. In the case of muon detectors, the main atmospheric influences on the measured cosmic ray intensity are related to variations of the atmospheric pressure and temperature (Sagisaka, 1986).

The barometric effect is observed as an anticorrelation between variations of the cosmic ray intensity and of the ground-level atmospheric pressure. This effect is more noticeable when low-pressure atmospheric events (such as tropical cyclones) pass the observation site. In these situations, we observe a clear cosmic ray intensity increase during the atmospheric pressure decrease. A good example of this kind of event is given in Figure 17 of Mendonça et al. (2016a). A simple explanation for this is an absorption through energy loss, dependent on the mass of atmosphere traversed. As this parameter can be related to the atmospheric pressure at a given altitude, we can say that the higher the atmospheric pressure, the higher the probability of a secondary cosmic ray particle being absorbed before reaching the ground. In addition to this absorption process, it is also expected that a pressure effect directly influences muon generation and decay in the atmosphere (Sagisaka, 1986).

The temperature effect, in turn, is also related to these two processes. For muons case, we expect a direct influence in their generation process and an indirect influence on their decay before reaching the ground. Muons are generated mainly by pion and kaon decay whose probability is directly proportional to the atmospheric temperature. The higher the temperature, the lower the atmospheric pion and kaon absorption that implies a higher generation rate of muons (Duperier, 1951). In this way, the higher the atmospheric temperature, the higher the muon production by this process (Maeda, 1960; Sagisaka, 1986; Dorman, 2004; Dmitrieva et al., 2011). However, due to atmospheric expansion occurring in the summer, muons have to travel a longer path before reaching ground-level detectors. Therefore, more low-energy muons are expected to decay before arriving at ground. On the contrary, during the winter, more muons are generated at relatively lower altitude allowing low-energy muons to reach the ground. In this way, the temperature effect can be separated in two parts: one called positive and other called negative. When we observe the cosmic ray intensity using a ground-level muon detector, the negative effect is predominant. Thus, we see a seasonal variation in antiphase with the temperature variation measured at the surface (Zazyan et al., 2015; Mendonça et al, 2016a; Mendonça et al, 2016b). On the other hand, the positive effect is more important on high-energy muon intensity observed by deep underground muon detectors whose data shows a seasonal variation in phase with the ground-level temperature (Adamson et al., 2010).

Many works have analyzed the pressure and temperature effects on the observed muon intensity through different methods (Ambrosio et al., 1997; Yanchukovskyet al., 2007; Adamson et al., 2010; Berkova et al., 2011; Dmitrieva et al., 2011; Tolkacheva et al., 2011; Braga et al., 2013; Mendonça et al., 2013; Rigozo, 2014; Zazyan et al., 2015; Mendonça et al., 2016; An et al., 2018; MaghrabI and Almutairi, 2018). By comparing several different methods, Mendonça et al. (2016a) found that the Mass-Weighted Method is the best for removing the temperature effect from the data recorded in the vertical channel of surface muon detectors. This method best reproduced the observed seasonal cosmic ray variation (related to atmospheric temperature changes). It also resulted in the highest correlation of the muon detector data corrected for the temperature effect with neutron monitor data, which are believed to be almost free of this effect.

The atmospheric pressure and temperature effects are related to the production, absorption and decay processes of secondary cosmic rays in the atmosphere. It is expected that the contribution from each of these processes depends on energy or rigidity of the secondary particles in the atmosphere. In the pressure case, we can say that the higher the energy of secondary particles the less they are absorbed by an atmospheric pressure increase, for example. In the temperature case, we expect that the negative temperature effect decreases as muon energy increases. In other words, low-energy muons are more affected by the atmospheric expansion in the summer. More detailed description about the dependence of barometric and temperature coefficients on secondary muon energy can be found in Sagisaka (1986) and Dorman (2004). As a first approximation, it is expected that higher energy primary particles generate higher energy secondary particles. Therefore, a good approximation for studying the energy dependence of atmospheric effects is to analyze how each effect is related to the geomagnetic cutoff rigidity ($R_C$) of the primary cosmic rays. In the analyses of the atmospheric effects in non-vertical directional channels, it is also necessary to consider the dependence on the zenith angle (Z) representing the path length in the atmosphere. Using neutron monitor data across different stations and by latitude surveys, past studies analyzed how barometric coefficient ($\beta$) is related to $R_C$. They found a clear anti-correlation: $\beta$ decreases as $R_C$ increases, i.e., the pressure effect becomes weaker with increasing $R_C$ (see section 6.9.1 of Dorman (2014) and references therein). As far we know, however, there are no reports analyzing the barometric effect dependency on cutoff rigidity using ground-level muon detector data. Likewise, there are no reports about the temperature effect behavior according to this parameter. In the case of the Global Muon Detector Network (GMDN), analysis of atmospheric effects on non-vertical field of views (i.e., at different zenith angles) were not performed yet. In this way, experimental studies of the relation between atmospheric coefficients on muon intensity and cutoff rigidity and zenith angle are still awaited to be explored.

In this work, we empirically analyzed the pressure and temperature effects on the Global Muon Detector Network (GMDN) data recorded between 2007 and 2016 and examine how both are related to primary particles cutoff rigidity. As described in Section 2.1, the GMDN observes muons arriving from various incident directions that are associated with different primary particles geomagnetic cutoff rigidities. Moreover, by using GMDN and temperature data described in Section 2.2, we can analyze the pressure and temperature effects on each directional channel. Thus, as shown in Section 3.1 and Section 3.2, we compare those coefficients with the average geomagnetic cutoff rigidity and the zenith angle associated with each GMDN directional channel. Finally, the summary of results and final remarks are presented in Sections 4 and 5, respectively.

## 2 – **Instrumentation**

The analyses presented in this paper are performed using cosmic ray and atmospheric data collected in the period between January 2007 and December 2016. More explicitly, we used: (I) the cosmic ray intensity observed in various directional channels of the Global Muon Detector Network (GMDN); (II) the ground-level atmospheric pressure measured at each detector site; and (III) atmospheric temperature profiles provided by the Global Data Assimilation System (GDAS) of the National Center for Environmental Prediction (NCEP).

### 2.1 - The Global Muon Detector Network (GMDN)

Four multidirectional muon detectors compose the GMDN. The oldest one is located at Nagoya (NGY) in Japan. It has been operating since early 1970s. The second oldest, which has been working since 1992, is at Hobart (HBT) in Australia. The remaining two are located at Sao Martinho da Serra (SMS) in Brazil and Kuwait City (KWT) in Kuwait. Both were installed in 2006. The prototype GMDN was formed in 2001 when a small SMS detector started operation in concert with NGY and HBT. The location and pictures of the four GMDN components are shown in Figure 1.

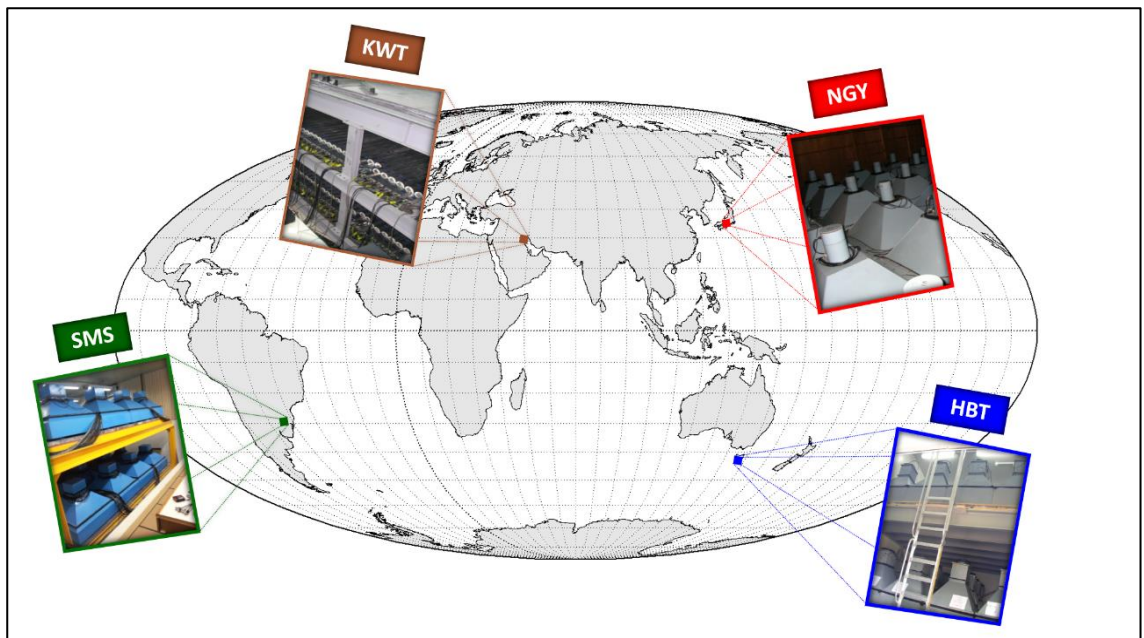

Figure 1 - Location and pictures of Sao Martinho da Serra (SMS), Kuwait (KWT), Nagoya (NGY) and Hobart (HBT) detectors.

As illustrated in Figure 2, NGY, HBT and SMS are formed by two horizontal layers of individual plastic scintillator detectors vertically separated by 1.73 m. Each of them is composed of: (I) a downward viewing photomultiplier tube at the top; and (II) a block of plastic scintillator with 1x1 m² area and 10 cm thickness located at the bottom. When a particle (red arrow) passes through the plastic scintillator, this material emits UV light (yellow symbol) that is converted to an electronic pulse by the photomultiplier, which is counted by an electronic system. A 5 cm thick lead layer is located below the upper individual detector layer to absorb low-energy background radiation. Only muons with energy higher than 300 MeV can pass through one upper individual detector and one lower layer detector to produce two-fold coincidence pulses. Comparing which upper and lower individual detectors observe a muon in coincidence, NGY, HBT and SMS electronic systems record the count rate of muons arriving from different individual incident directions.

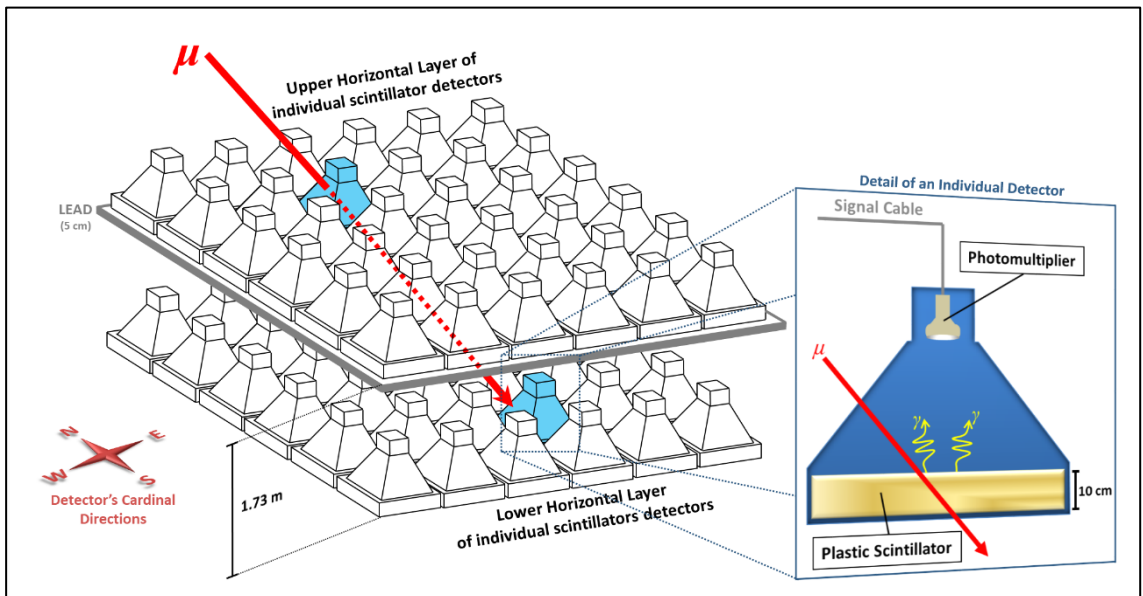

Figure 2 - Sketch of the Nagoya muon detector. Hobart and Sao Martinho da Serra detectors are similar except for the individual scintillators detectors number on each horizontal layer. There are 4x4 in the first and 4x9 individual detectors in the second. The red cross in the left shows the detector's cardinal directions that is aligned to the geographic ones in NGY and SMS. Since December 2010, after its enlargement, HBT was rotated about 28° clockwise.

As shown in Figure 3, the KWT is composed of four horizontal layers of cylindrical proportional counter tubes. Each tube is 5 m long and 10 cm in diameter with a 50 µm thick tungsten anode wire along the cylinder axis. In two layers tubes are aligned in the X direction while in the other two layers, tubes are orthogonally aligned along the Y direction. The X layers are rotated 32.8° anticlockwise from the north geographic direction. The detector consists of two pairs of X and Y tubes layers vertically separated by 80 cm. A 5 cm thick lead layer is located above these two pairs to absorb low-energy background radiation. In a similar way to the other GMDN detectors, KWT electronic system can monitor the cosmic ray intensity in various directional channels by identifying which tube in each layer is traversed by a muon.

In the analysis period of this work, i.e. between January 2007 and December 2016, detection areas of GMDN were expanded in several steps, except NGY which has had the same detection area of 36 m² since 1970. Until November 2010, HBT had a 9 m² detection area. After that it was enlarged to 16 m². The detection area of SMS also increased from 28 m² to 32 m² in September 2012 and from 32 m² to 36 m² in July 2016. Finally, KWT detection area was increased from 9 m² to 21.5 m² in April 2015 and to 25 m² in April 2016.

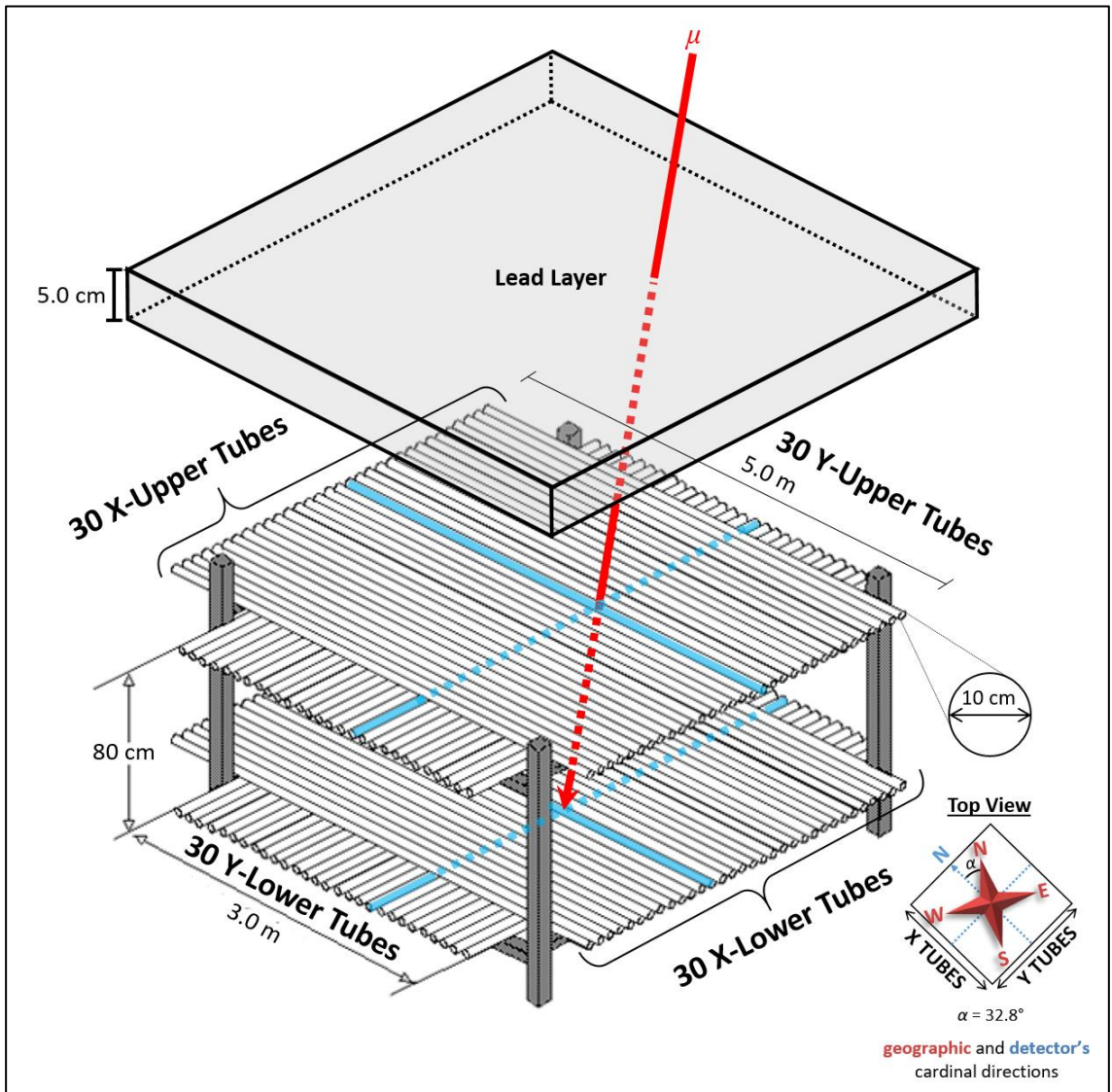

Figure 3 - Sketch showing the Kuwait muon detector. After April 2015, the number of X and Y tubes increased to 43 and 50, respectively.

Since 2007, GMDN detectors started recording data with a new electronic system described by Yassue et al. (2003). NGY, SMS and HBT started monitoring cosmic ray intensities in 121, 105 and 49 directional channels, respectively. KWT has been monitoring 529 directional channels since its installation, but many of them have a large statistical error due to their small detection area before 2015. We therefore analyzed data from KWT by grouping 3x3 neighboring channels to decrease this error. Thus, the field of view of this detector composes a 7x7 matrix of grouped directional channels each with larger detection area, instead of the original 23x23 matrix of channels with smaller area.

Figure 4 shows the color-coded map of hourly muon count rate 1σ error in % calculated for each GMDN directional channel. Each panel in this figure represents the field of view of each detector in December 2008. Small squares inside each panel represent the directional channels with their color indicating the calculated count rate error. The x and y coordinates associated with each square define each channel's field of view direction according to the relative position between the upper and lower individual detectors that compose it. While the x-axis is aligned with the detector's cardinal west-east direction, the y-axis is aligned with the detector's north-south direction. For example, the directional channel [-1,3] shows the coincidence detections made by an upper individual detector located "one individual detector" to the west and "three" to the north from the lower one whose coincidence detection was made, which is the case shown by the red arrow on Figure 2. As already shown on caption of this figure, NGY and SMS cardinal directions are aligned with the geographic ones, while HBT cardinal directions are rotated clockwise about 28° since December 2010. For KWT, the position of the directional channel is given by the relative position between the upper and lower pair of orthogonal tubes instead of individual scintillation detector. Moreover, the central square, which is given by the coordinates [0,0], represents the vertical directional channel that takes into account only particles that are observed by one upper individual detector located exactly above the lower individual detector that observed this particle in coincidence.

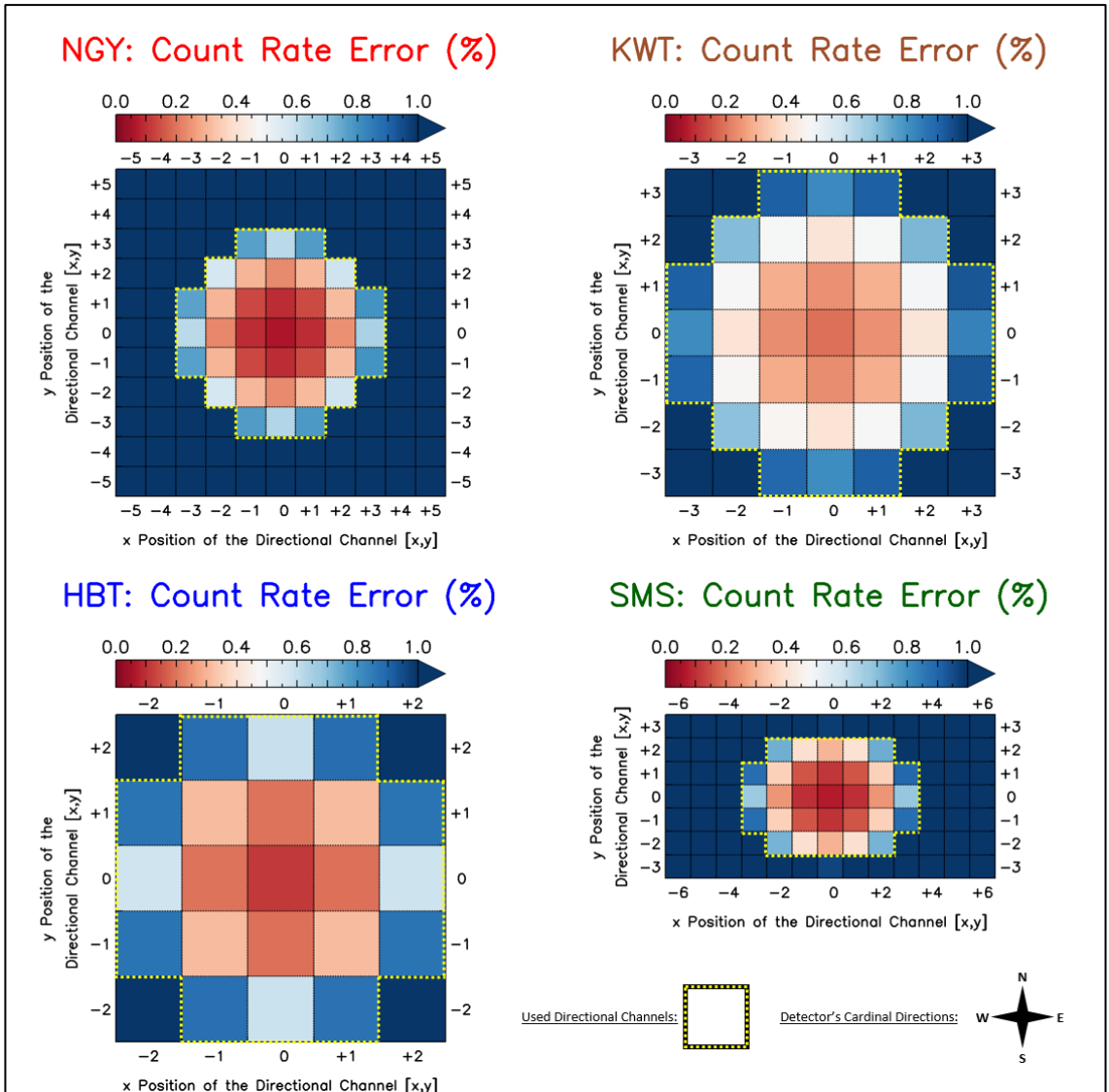

Figure 4 - The 1σ Count Rate Error obtained for each GMDN directional channel. The four boxes, from top to bottom and left to right, show respectively Nagoya (NGY), Kuwait (KWT), Hobart (HBT) and Sao Martinho da Serra (SMS) data. The colored squares inside each detector box represent the relative standard deviation in percentage calculated considering hourly data recorded in December 31, 2008 according the following equation: $100/\sqrt{\langle C_r \rangle}$, with $\langle C_r \rangle$ denoting the average count rate in this period. The x and y position indicates the channel [x,y] field of view direction with respect to the origin. The dashed yellow line delimits the directional channels with lower error that were selected for the analysis shown in this work.

It is clear in Figure 4 that very inclined directional channels are associated with larger count rate errors. This is partly due to the small number of upper and lower individual detector combinations that can be used to compose them, i.e., due to their smaller detection area. Because of that, we do not use them for the analysis in this work and use only the directional channels inside the dashed yellow lines indicated in each panel, which includes only the directional channels with count rate error below 1%. Strictly speaking, HBT, KWT and SMS detectors did not present the same count rate error in the whole period of analysis due to the already mentioned detection area enlargements. As all enlargements were done after 2008, the count error of these detectors are smaller than those shown in Figure 4 in the present-day.

We consider the interaction of primary cosmic ray particles with the geomagnetic field using the Smart et al. (2000) model. More information and computational code can be found at the Community Coordinated Modeling Center (CCMC) webpage (https://ccmc.gsfc.nasa.gov/modelweb/sun/cutoff.html). Using this model, we calculated the effective geomagnetic cutoff rigidity ($R_C$) expected for each GMDN directional channel $[x, y]$ used in this work. In this calculation, we consider only a proton's orbit arriving with the zenith and azimuth angles corresponding to those that result in the highest effective detection area. For example, only muons with 0° zenith angle can hit the whole area of the upper and lower individual detectors and be detected in the vertical directional channel. As particle zenith angle increases, muons can hit only a fraction of the individual detector's detection area in order to form this directional channel. In this way, although the vertical directional channel observes particles arriving with zenith angles from 0° to 39°, most of the muons detected in this channel arrive with zenith angles close to zero. Similarly, the [0,2] directional channel, for instance, observes muons with zenith angle between 30-61 degrees, but the highest effective detection area can be only achieved when considering 49°. In the azimuth case, this condition will be attained only when considering 0° in the angular range between -27° and 27°. In this way, we can use the following equations:

$$Z = tan^{-1}\left(\frac{\sqrt{x^2 + y^2}}{h}\right) \quad (1)$$

$$A = tan^{-1}\left(\frac{x}{y}\right) \quad (2)$$

where $Z$ and $A$ are respectively the zenith and azimuth angles of the directional channel given by the position at $[x, y]$ coordinates in Figure 4 and $h$ is the vertical distance between the upper and lower detector layers (1.73 m for NGY, HBT and SMS). For KWT, $h$ is 0.8 m and $[x, y]$ should be multiplied by 0.3 before using them. The values of zenith angle of all GMDN channels used in this work are provided in the Supporting Information.

Besides $Z$ and $A$ values found for each directional channel, we calculate annual values of $R_C$ taking into account Definitive Geomagnetic Reference Field (DGRF) coefficients (for 2000, 2005 and 2010) and IGRF-12 coefficients (for 2015) to calculate the annual main geomagnetic field. More information about DGRF and IGRF coefficients and models can be found at the International Association of Geomagnetism & Aeronomy (IAGA) webpage (https://www.ngdc.noaa.gov/IAGA/vmod/index.html). For the HBT detector, we also consider its rotation (azimuth change) after 2010. Finally, from the 10 annual values obtained from 2007 and 2016, we calculate the average geomagnetic cutoff rigidity for each GMDN directional channel $\langle R_C \rangle$, whose values are shown in Figure 5. In this figure, we can see that $\langle R_C \rangle$ for HBT, which is located closer to the South Geomagnetic Pole, is about 10 GV lower than those for other detectors. Moreover, it is also seen that directional channels observing particles coming from the East (with x > 0) have higher $\langle R_C \rangle$ than those monitoring particles incident from the West (with x < 0). This is related to the well-known geomagnetic East-West effect, which implies fewer low-energy cosmic rays (mostly protons with positive charge) can arrive at Earth's surface from eastern direction than from western direction. In this way, we observed fewer low-energy particles arriving from this direction. In Figure 5, we can clearly see this effect particularly in $\langle R_C \rangle$ calculated for NGY and SMS. A small East-West Effect is also present in HBT data. For example, while an East directional channel at [+2,0] position has $\langle R_C \rangle$ about 2.1±0.1 GV, we found 1.8±0.1 GV for the corresponding west channel at [-2,0]. The difference between $\langle R_C \rangle$ values at [+2,0] and [-2,0] is, on the other hand, larger than 11 GV at NGY.

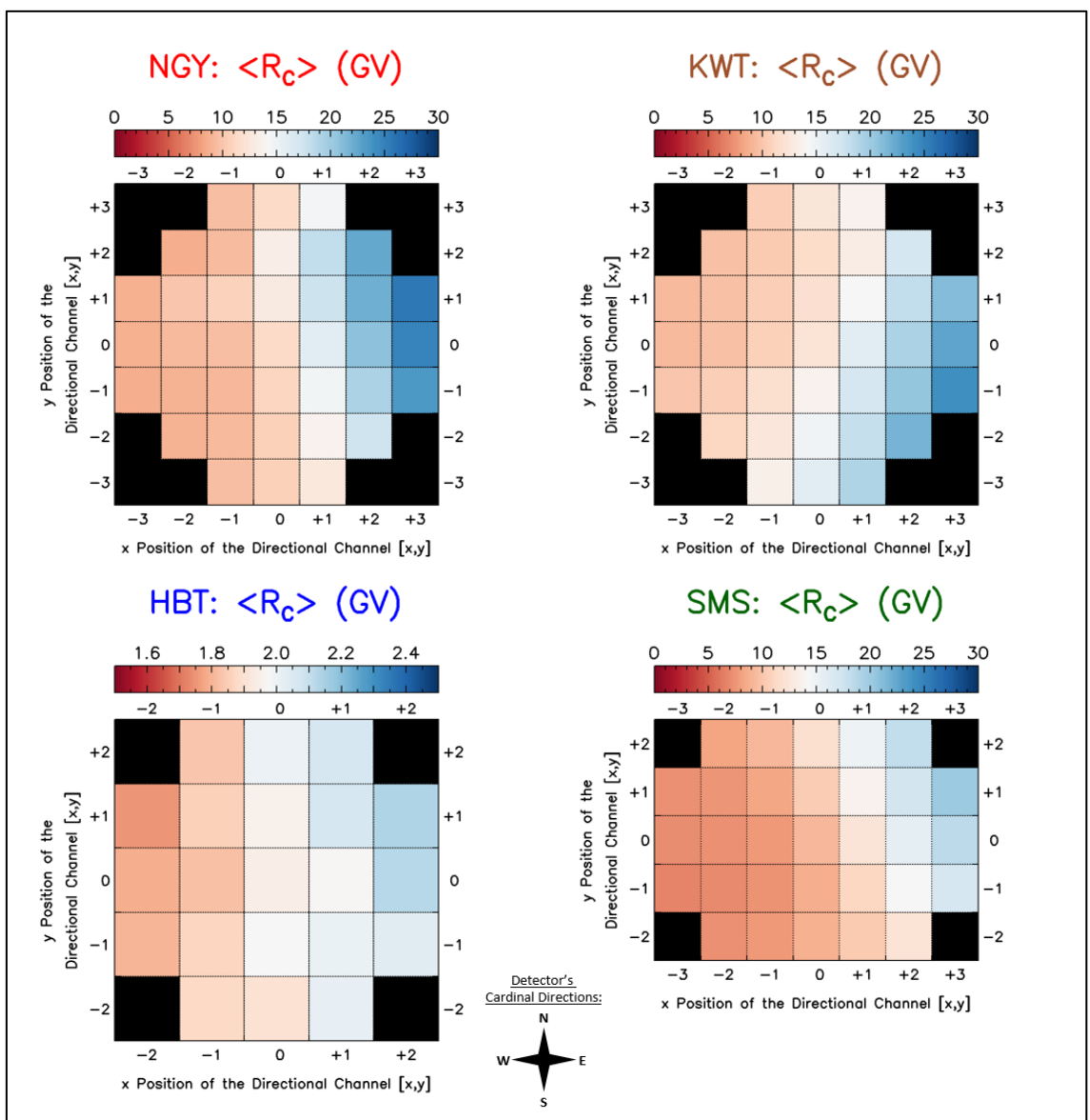

Figure 5 - Average Effective Geomagnetic Cutoff Rigidity obtained for the Global Muon Detector Network. The four boxes, from top to bottom and left to right show, respectively, Nagoya (NGY), Kuwait (KWT), Hobart (HBT) and Sao Martinho da Serra (SMS) data. The colored squares inside each detector box represent the average values obtained for each directional channel of this detector. The x and y position indicate the direction of the channel's field of view with respect to the origin in a way that channel [0,0] is the one that observes particles vertically arriving at the detector. The black squares indicate directional channels not used in this work. Note that HBT does not have the same color table range as used for the remaining detectors. The values used to produce this figure are available in the Supporting Information tables.

As previously stated, the main purpose of this paper is to study the relation of pressure and temperature effects with geomagnetic cutoff rigidity associated with different fields of view. Thus, using GMDN data with count rate errors less than 1 %, as shown in Figure 4, we compare the relation between $\langle R_C \rangle$ values shown in Figure 5 and the coefficients found for these effects when analyzing each directional channel data individually.

2.2 - Atmospheric Pressure and Temperature Data

For obtaining these pressure and temperature effects coefficients, we use the ground-level atmospheric pressure and the mass-weighted atmospheric temperature at four GMDN sites, together with the observed muon data. While we use the ground-level pressure measured at each site, we derive the mass-weighted atmospheric temperature from the global meteorological data provided by the United States National Oceanic and Atmospheric Administration (NOAA). In more detail, we use hourly atmospheric pressure measured by a piezoelectric type pressure sensor installed at each GMDN site and the atmospheric temperature profiles recorded by the Global Data Assimilation System (GDAS) maintained by the NOAA's Air Resources Laboratory (ARL) webpage (https://www.ready.noaa.gov/gdas1.php).

The GDAS system compiles many kinds of meteorological observations (such as balloons, ground and satellite measurements) each as a function of the 3D location (geographic longitude, latitude and altitude) on Earth. In this work, we use 3-hourly atmospheric temperature profiles obtained for every 1° by 1° surface grid around each GMDN site and scaled in 24 fixed atmospheric pressure levels. Following Mendonça et al. (2016a) results, we adopt the Mass-Weighted Method to describe the temperature effect. In this way, we compile the atmospheric temperature profiles in single variable as shown below:

$$T^{MSS}(t) = \sum_{l=0}^{23} \frac{x_l(t) - x_{l+1}(t)}{x_0(t)} * T_l(t) \qquad (3)$$

where $T^{MSS}(t)$ is the atmospheric temperature weighted by air mass at time $t$; $x_l(t)$ and $T_l(t)$ are, respectively, the atmospheric depth and temperature for the level $l$ observed at the same time. While $l = 0$ corresponds to the GDAS closest to ground pressure level (1000 hPa), $l$ = 23 corresponds to the highest altitude (20 hPa, about 26.5 km altitude). We assume that the level 24 is equivalent to top of the atmosphere and that $x_{24}(t)$ = 0.

**3 - Analysis and Results**

In this section, we describe the analysis of the atmospheric effects on secondary cosmic ray muons arriving at observation site with different directions of view. First, we present the study about the barometric effect and then about the temperature effect.

3.1 - Pressure Effect Analysis

Considering barometric effect theory, see e.g., Sagisaka (1986) or Appendix A of Mendonça et al. (2016a), we define the atmospheric pressure effect on the muon count rate, as:

$$ln\left[\frac{I_{[x,y]}(t)}{\langle I_{[x,y]} \rangle}\right] * 100 \% = \beta_{[x,y]} * [P(t) - \langle P \rangle] \qquad (4)$$

where $I_{[x,y]}(t)$ is the cosmic ray count rate observed in directional channel at the $[x, y]$ position at time $t$; $P(t)$, given in hPa, is the ground atmospheric pressure measured on the detector site at the same time; $\langle I_{[x,y]} \rangle$ and $\langle P \rangle$ are both reference values (in this work, the mean values of $I_{[x,y]}(t)$ and $P(t)$ obtained in the period of analysis, respectively); and $\beta_{[x,y]}$ is the barometric coefficient in %/hPa representing how much the pressure effect influences the observed cosmic ray intensity. Hereinafter, the $\beta_{[x,y]}$ will only be denoted by $\beta$.

In this work, we obtained $\beta$ from the linear regression between experimental hourly cosmic-ray and atmospheric pressure data. To do so, we need to choose a period of analysis where other (solar, interplanetary, geomagnetic and atmospheric) effects are not present or have little influence on the cosmic ray intensity observed at ground. Considering this, we calculated the barometric coefficient in short periods (one month) to avoid long-term variations not related to the pressure effect and only consider months when the pressure effect is significant. In more detail, for each detector we only selected the months where the absolute value of the linear Pearson correlation coefficient between hourly vertical directional channel and pressure data are higher than 0.7. In this way, out of 120 months comprising the total analysis period, 116 months for HBT, 78 months for NGY, 55 months for SMS and 4 months for KWT were used. Finally, we obtained an average barometric coefficient considering only values found in these good correlation periods.

Figure 6 shows the average barometric coefficient obtained for each GMDN directional channel ($\langle\beta\rangle$) used in this work. We can notice that values of $\langle\beta\rangle$ found for HBT are lower than those found for other detectors. While they are around -0.16 and -0.15 %/hPa, the barometric coefficients are between -0.15 and -0.10 %/hPa on NGY, KWT and SMS. Moreover, it is also possible to see an East-West asymmetry in $\langle\beta\rangle$. Particularly in HBT, NGY and SMS data, the right side (x > 0) of detectors field of view has a lighter color (closer to zero) than the left side. In KWT data, we notice better a northwest-southeast asymmetry, i.e., the lower-right channels are closer to zero than the upper-left channels. This behavior, which is similar to that observed in the cutoff rigidity values (Figure 5), is related to the KWT detector inclination with respect to the geographic cardinal directions. As shown in the right corner of Figure 3, the upper-left and lower-right channels in KWT are respectively closer to the west and east geographic directions (defined by the red cross). Lastly, we can notice that although this East-West asymmetry of $\langle\beta\rangle$ is also present for the HBT detector, it is very small and within the margin of errors. For example, $\langle\beta\rangle$ found for the east directional channel [+2,0] is -0.151 ± 0.006 %/hPa while the value found for the channel [-2,0] (i.e., the equivalent channel in the west direction) is -0.154 ± 0.006 %/hPa. Besides this east-west asymmetry, in HBT, we also notice a significant variation according to the zenith angle of each directional channel field of view. The less inclined (central) channels are darker than the more inclined ones (located at borders). In other words, $\langle\beta\rangle$ tends to be closer to zero as the channels inclinations increases with larger $x^2 + y^2$. Thus, the pressure effect seems to be weaker for very inclined directional channels.

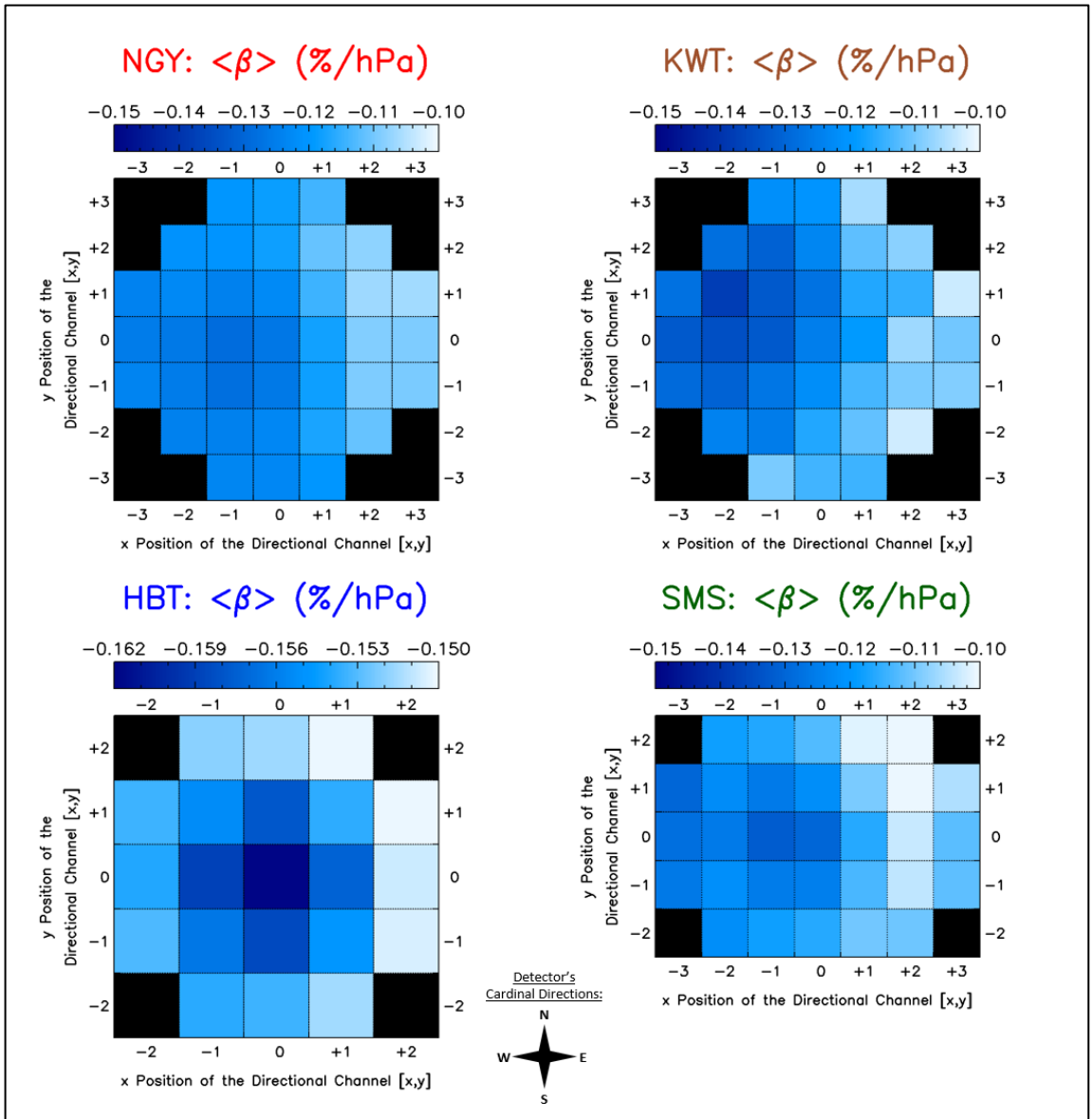

Figure 6 - Average Barometric coefficients found for the Global Muon Detector Network. The four boxes, from top to bottom and left to right, show respectively Nagoya (NGY), Kuwait (KWT), Hobart (HBT) and Sao Martinho da Serra (SMS) data. The colored squares inside each detector's box represent the average values obtained for each directional channel of this detector. The x and y position indicate the direction of channel's field of view with respect to the origin in a way that channel [0,0] is the one that observes particles vertically arriving at the detector. The black squares indicate directional channels not used in this work. The values used to produce this figure are available in the Supporting Information tables. Note that HBT does not have the same color table range as used for the remaining detectors.

The East-West asymmetry of average barometric coefficients in Figure 6 resembles that seen in averaged geomagnetic cutoff rigidity in Figure 5. A better comparison between both is demonstrated in Figure 7 where $\langle \beta \rangle$ is shown as a function of $\langle R_C \rangle$. The vertical error bar of each point is deduced from the uncertainty of monthly barometric coefficients used in this average. The horizontal error bar, which is smaller than the symbol size in most cases, is deduced from the standard deviation of annual values used to calculate $\langle R_C \rangle$. In this figure, we can clearly see that $\langle \beta \rangle$ and $\langle R_C \rangle$ seem to have a natural logarithm relation that can be summarized by the black curve. This curve is obtained by a linear regression between $\langle \beta \rangle$ and $ln(\langle R_C \rangle)$. In this case, the Pearson Correlation Coefficient (PPC) is higher than 0.9. However, it would be very difficult to notice this logarithm relation without HBT data (blue diamonds) since other detectors points can be fitted by a linear expression between $\langle \beta \rangle$ and $\langle R_C \rangle$. As we already discussed, HBT barometric coefficients and cutoff rigidities found for HBT are smaller than those found for KWT, NGY and SMS. In this way, when looking all detectors together, we can notice a fast decrease of $\langle\beta\rangle$ with the decrease of $\langle R_C\rangle$ indicating a logarithm relation between both.

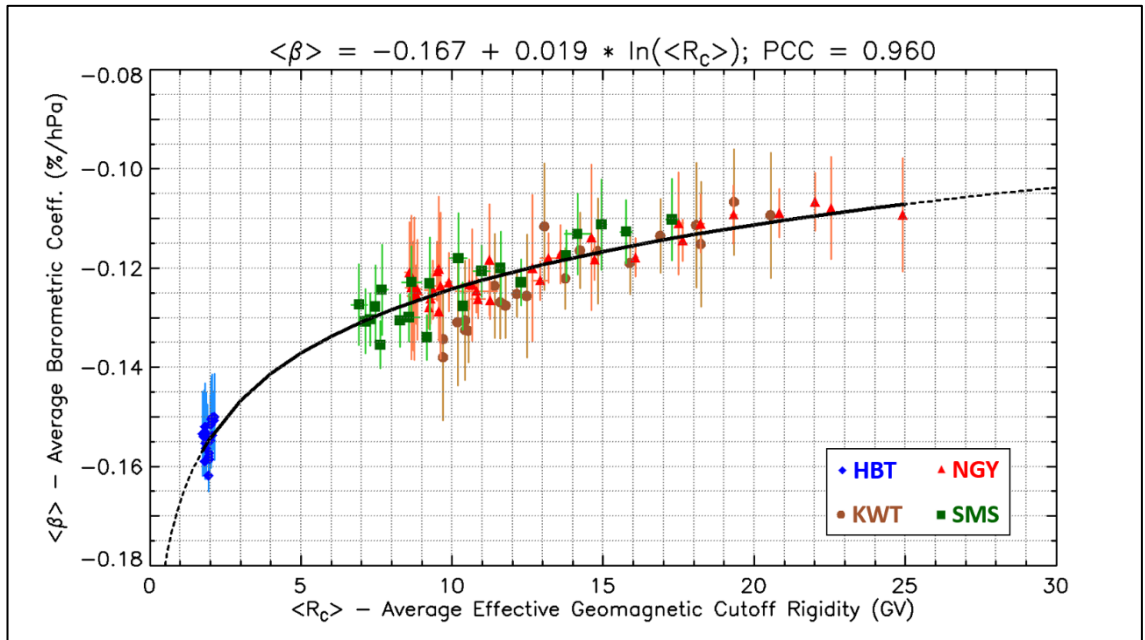

Figure 7 - Correlation between the average values of barometric coefficient and effective geomagnetic cutoff rigidity associated with each GMDN directional channel. The blue diamonds, brown circles, red triangles and green squares show data from Hobart (HBT), Kuwait (KWT), Nagoya (NGY) and Sao Martinho da Serra (SMS) detectors, respectively. The vertical and horizontal bars along with each point represent $\langle\beta\rangle$ and $\langle R_C\rangle$ uncertainties, which are not considered in the black curve calculation. The equation that defines this curve and the Pearson Correlation Coefficient (PPC) value found in this case are shown in the top. The values used to produce this figure are available in the Supporting Information tables.

The barometric coefficients of some channels presented a high uncertainty. In order to better analyze the $\langle\beta\rangle$ dependency on cutoff rigidity, we chose to focus only on cases where the pressure coefficient error is lower than 0.01 %/hPa. As shown in the top box of Figure 8, in this situation the total number of points decreases. In this case, we found that the Pearson Correlation Coefficient of the linear relation between $\langle\beta\rangle$ and $ln(\langle R_C\rangle)$ changes from 0.960 (Figure 7) to 0.974 (Figure 8-A). In addition, we can see a good "linear" alignment when analyzing NGY (red triangles), KWT (brown circles) or SMS (green squares) points alone. On the other hand, we cannot see that if we consider only the HBT data (blue diamonds). While the natural logarithmic between the pressure effect coefficients and cutoff rigidities presents a correlation coefficient about 0.9 in the first case, it is about 0.3 in the second. Considering the barometric coefficient behavior according to the directional channel field of view inclination on HBT data, which was discussed in Figure 6 analysis, we assume that $\langle\beta\rangle$ found for a directional channel may be also directly related to the zenith angle (Z) of this channel. Since the pressure effect coefficients tends towards zero as the channel inclination increases ($Z \to 90°$), we decide to analyze the relation between $\langle\beta\rangle$ and $\langle R_C\rangle/cos(Z)$. As we can see in the bottom box of Figure 8, after that, the HBT points set (blue diamonds) shows a very clear "linear" distribution. At the same time, NGY and SMS data (red triangles and green squares, respectively) also show good alignment on the fitted (black) curve. The correlation coefficient between the barometric coefficient and the natural logarithmic of the cutoff rigidity divided by zenith angle cosine is 0.994 when considering the three detectors data together. Thus, we can say that $\langle\beta\rangle$ presents better correlation with $ln[\langle R_C\rangle/cos(Z)]$ than with $ln(\langle R_C\rangle)$.

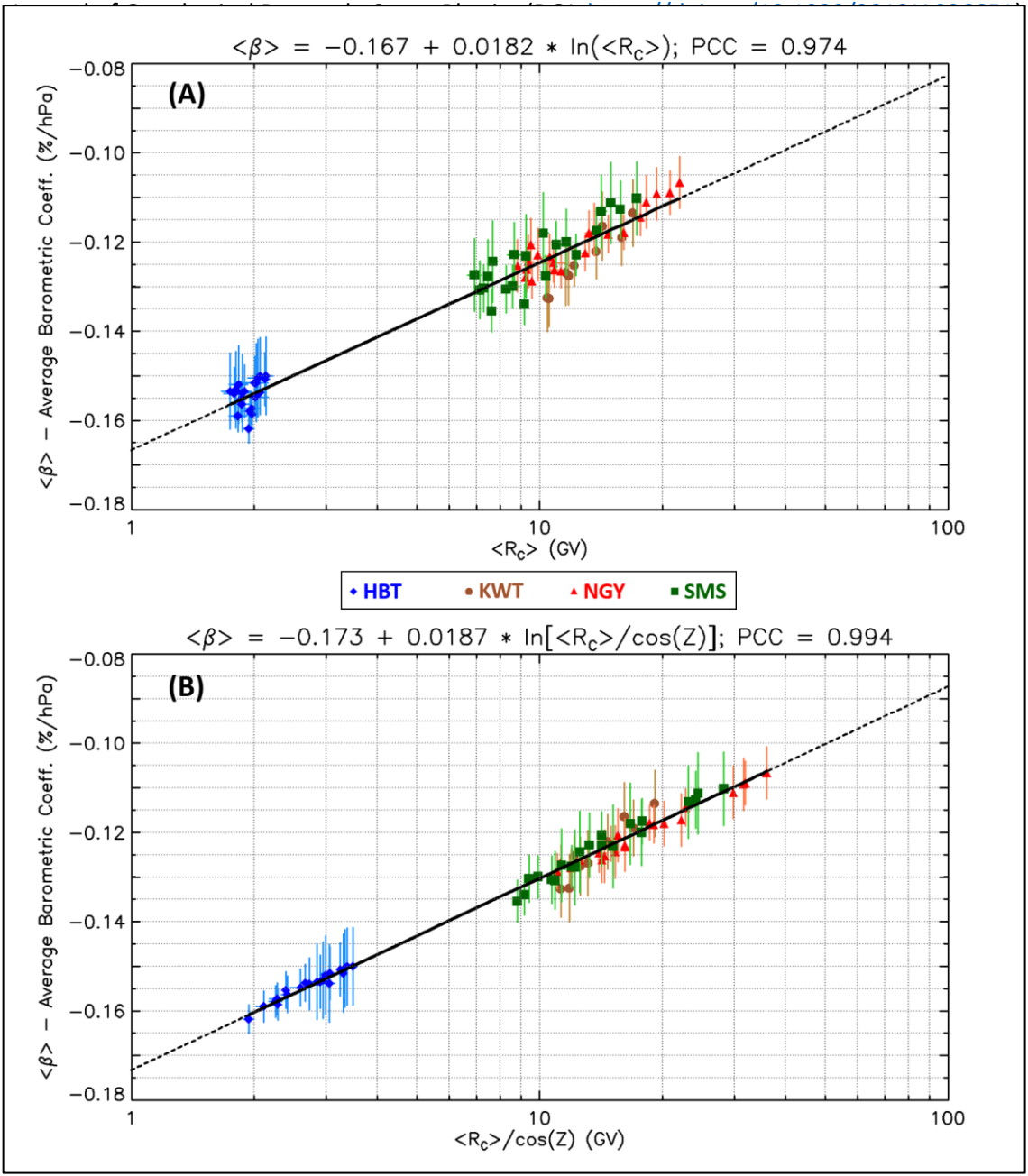

Figure 8 - Correlation between the average values of barometric coefficient ($\langle\beta\rangle$) and (A) the average geomagnetic cutoff rigidity ($\langle R_C \rangle$) and (B) the quotient between $\langle R_C \rangle$ and the zenith angle ($Z$) cosine found for each GMDN directional channel. In both boxes, blue diamonds, brown circles, red triangles and green squares show data from Hobart (HBT), Kuwait (KWT), Nagoya (NGY) and Sao Martinho da Serra (SMS) detectors, respectively. The vertical and horizontal bars along with each point represent $\langle\beta\rangle$ and $\langle R_C \rangle$ uncertainties, which are not considered in the black curve calculation. The equation that defines this curve and used to calculate the Pearson Correlation Coefficient (PPC) value is shown in the top of each box. The values used to produce this figure are available in the Supporting Information tables.

### 3.2 - Temperature Effect Analysis

In this work, as already discussed, we use the Mass-Weighted Method to describe the temperature effect on ground muon detectors data. This method considers temperature variation in the entire atmosphere through a single coefficient that can be empirically calculated according to the following equation:

$$\frac{I^{CP}_{[x,y]}(t) - \langle I^{CP}_{[x,y]} \rangle}{\langle I^{CP}_{[x,y]} \rangle} * 100\,\% = \alpha^{MSS}_{[x,y]} * [T^{MSS}(t) - \langle T^{MSS} \rangle] \qquad (5)$$

where $I^{CP}_{[x,y]}(t)$ is the muon count rate corrected for the pressure effect using $\langle\beta\rangle$ values shown in Figure 6; $T^{MSS}(t)$ is the mass weighted atmospheric temperature in K; $\langle I^{CP}_{[x,y]}\rangle$ and $\langle T^{MSS}\rangle$ are the mean values of $I^{CP}_{[x,y]}(t)$ and $T^{MSS}(t)$ in the period of analysis; and $\alpha^{MSS}_{[x,y]}$ is the mass weighted temperature coefficient in %/K. Hereinafter, the $\alpha^{MSS}_{[x,y]}$ will only be denoted as $\alpha^{MSS}$.

We calculate $\alpha^{MSS}$ through a linear regression between $I^{CP}_{[x,y]}(t)$ and $T^{MSS}(t)$ over a one-year period. Since the dominant variation of the temperature is the seasonal variation, a one-year time window covers its maximum and minimum periods (i.e. a whole cycle). In this way, we can calculate $\alpha^{MSS}$ from one year of data, avoiding strong influences of long-term modulation of cosmic rays related to solar activity.

Thus, using GMDN and GDAS data recorded between January 2007 and December 2016, we first obtain ten annual values of $\alpha^{MSS}$. Similarly to what we have done in the barometric effect analysis, we calculate an average of the temperature effect using only $\alpha^{MSS}$ obtained in periods when there is a good correlation between $I^{CP}_{[0,0]}(t)$ and $T^{MSS}(t)$. Explaining in more detail, if the PCC is below 0.7 in a chosen one-year period and detector, we discard $\alpha^{MSS}$ obtained for all directional channels of this detector in this year. In this way, from the total ten periods which we have, only eight and five years are used for deriving $\alpha^{MSS}$ for HBT and SMS, respectively, while no year is discarded for deriving $\alpha^{MSS}$ for NGY and KWT.

Figure 9 shows mass weighted temperature coefficients average values found for each GMDN directional channel considering only the selected years. In this figure, it is seen that $\langle\alpha^{MSS}\rangle$ for the four detectors are quite different from each other. Temperature coefficients found for NGY and HBT are below -0.23 %/K, while $\langle\alpha^{MSS}\rangle$ for KWT are mainly between -0.25 and -0.22 %/K. Finally, they are higher than -0.22 %/K for SMS. In this way, we can roughly say that the temperature effect seems to be stronger (with larger negative coefficient) for HBT and NGY than for KWT and is weakest for SMS.

In addition to the difference from one detector to another, Figure 9 also shows how $\langle\alpha^{MSS}\rangle$ varies according to the [x,y] coordinate of the directional channel. There seems to be a kind of east-west asymmetry and a dependence on inclination of each directional channel. As with the pressure effect analysis, we can see, mainly in HBT data, a dependency related to inclination of the directional channel field of view. On the other hand, contrary to pressure coefficients, $\langle\alpha^{MSS}\rangle$ decreases as the directional channel inclination increases. It is easy to see in the HBT box that the border squares have a darker color than the centered (less inclined) ones. It is also possible to notice that this change occurs symmetrically in a way that channels whose x and y positions applied in the expression x²+y² results in the same value, tending to present similar $\langle\alpha^{MSS}\rangle$. For example, $\langle\alpha^{MSS}\rangle$ found for HBT are around -0.25 %/K for pixels with x²+y²=1 (channels [1,0], [-1,0], [0,1] and [0,-1]), while they are around -0.26 %/K for all pixels with x²+y²= 2 (for the channels [1,1], [1,-1], [-1,1] and [-1,-1]). This increase of $\langle\alpha^{MSS}\rangle$ with the channel inclination (zenith angle) is harder to see for KWT, NGY and SMS probably because it is obscured by the strong east-west asymmetry present in the temperature coefficients found for these detectors.

A clear east-west asymmetry of $\langle\alpha^{MSS}\rangle$ is seen for NGY and SMS in Figure 9. For both detectors, the coefficients associated with the directional channels pointing to east are closer to zero than that found on the equivalent channels pointing to west. In a similar way, besides the strong variation with the inclination, we can also see that HBT channels on the right side tend to present a lighter color than their equivalent channel on the left side. For the KWT detector, we can also observe an asymmetry that is more visible when comparing the directional channels looking to Northwest with those looking to Southeast. While the temperature coefficients of the first group are lower than -0.24 %/K (have a darker color), those of the second group are higher than that value (have a lighter color). As previously mentioned, the KWT detector, different from NGY and SMS, is not exactly aligned to the geographic directions. As shown in the right corner of Figure 3, the upper-left and lower-right channels in KWT are respectively closer to the west and east geographic directions (defined by the red cross). As shown in Figure 5, an asymmetry between these two regions can be also seen on average cutoff rigidities found for KWT detector. We can also see in this figure that $\langle R_C \rangle$ found for NGY and SMS shows a pronounced east-west difference like that observed in $\langle \alpha^{MSS} \rangle$ for these detectors. For HBT, the east-west asymmetry of $\langle R_C \rangle$ in Figure 5 looks different from that seen in Figure 9 due to the strong variation of $\langle \alpha^{MSS} \rangle$ with the directional channel inclination.

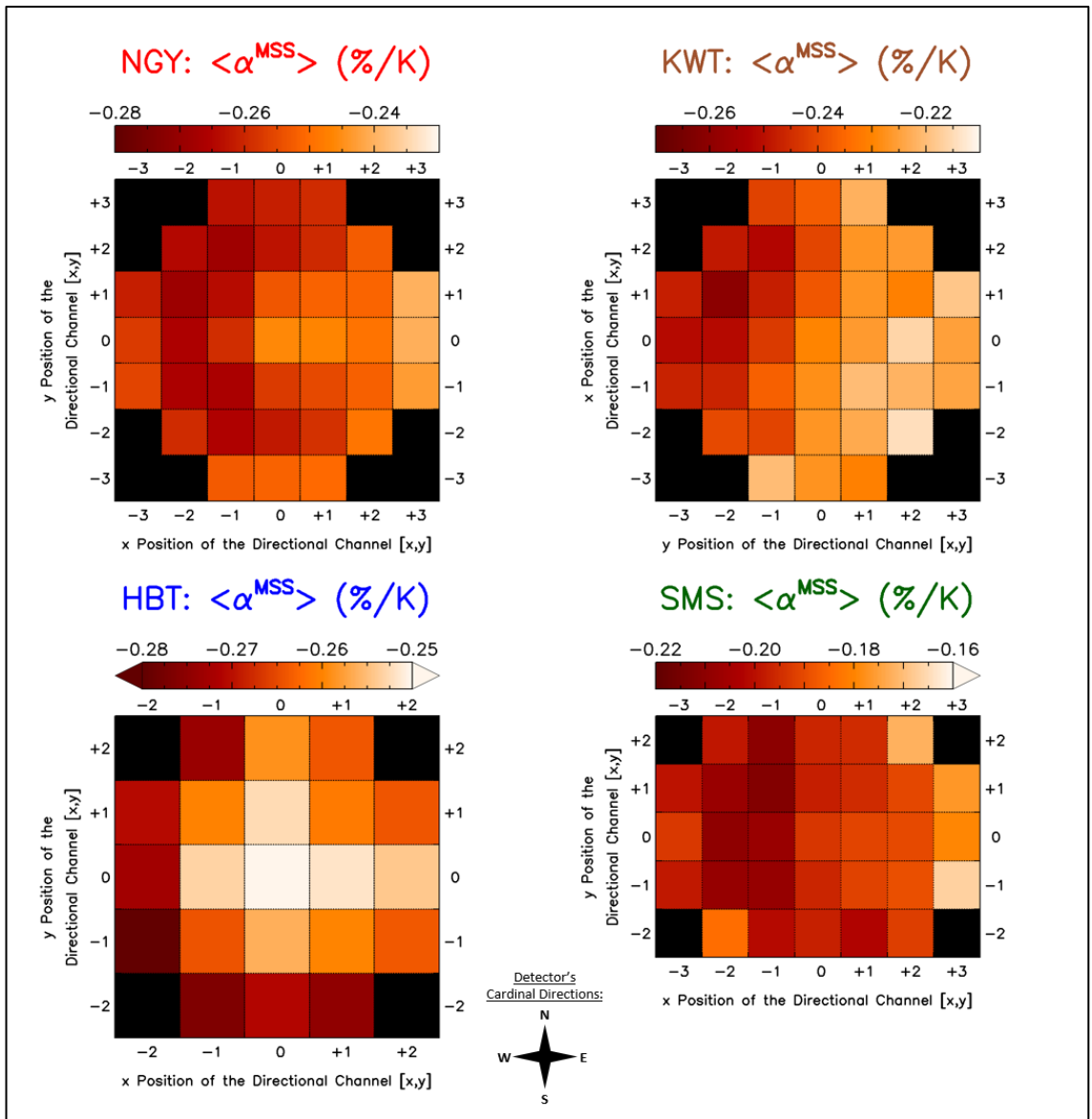

Figure 9 - Mass weighted temperature coefficients found for the Global Muon Detector Network. The four boxes, from top to bottom and left to right, show respectively Nagoya (NGY), Kuwait (KWT), Hobart (HBT) and Sao Martinho da Serra (SMS) data. The colored squares inside each detector's box represent the average values obtained for each directional channel of this detector. The x and y position indicates the direction of channel's field of view with respect to the origin in a way that channel [0,0] is the one that observes particles vertically arriving at the detector. The black squares indicate directional channels not used in this work. The values used to produce this figure are available in the Supporting Information tables. Note that each detector has a different color table range.

Figure 10 shows a better comparison between mass-weighted temperature coefficients and the geomagnetic cutoff rigidity associated with each GMDN directional channel. In this figure, only $\langle \alpha^{MSS} \rangle$ points that have errors lower than 0.015%/K are shown and it is possible to see that there are multiple relations between $\langle \alpha^{MSS} \rangle$ and $\langle R_C \rangle$. As shown by the dashed curves, these two variables present a different linear relationship for each detector. We can see that the slopes of the linear relations found for SMS (green), NGY (red) and KWT (brown) are quite similar. It seems that the values of $\langle \alpha^{MSS} \rangle$ for these detectors are separated from each other by an offset which seems to be independent of $\langle R_C \rangle$ and $Z$. On the other hand, the linear curve for HBT (blue) has a slope very different from other detectors. As already discussed, HBT has a unique feature in the GMDN. Its directional channels are associated with a cutoff rigidity range that is about 10 GV lower than the range covered by KWT, NGY and SMS detectors (see Figure 5). Moreover, only $\langle \alpha^{MSS} \rangle$ for HBT shows a clear variation with zenith angle instead of the east-west asymmetry (see Figure 9).

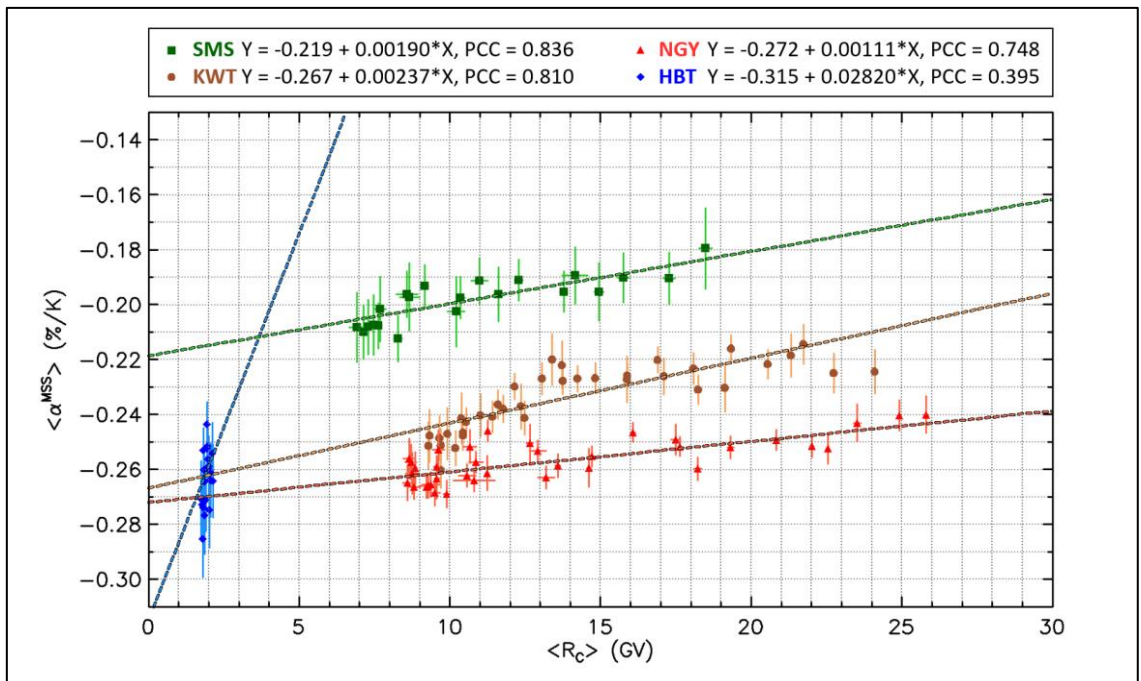

Figure 10 - Correlation between the average values of mass weighted temperature coefficients ($\langle \alpha^{MSS} \rangle$) and the average geomagnetic cutoff rigidity ($\langle R_C \rangle$) found for the GMDN. The blue diamonds, brown circles, red triangles and green squares show data from Hobart (HBT), Kuwait (KWT), Nagoya (NGY) and Sao Martinho da Serra (SMS) detectors, respectively. The dashed blue, brown, red and green lines show linear correlations found using data with the same color. The equations that define each dashed line and their respective Pearson Correlation Coefficients (PPC) are displayed at the top of the plot. The vertical and horizontal bars along with each point represent the uncertainties associated with $\langle \alpha^{MSS} \rangle$ and $\langle R_C \rangle$, which are not considered in dashed curves calculation. Only $\langle \alpha^{MSS} \rangle$ points that have errors lower than 0.015%/K are shown. The values used to produce this figure are available in the Supporting Information tables.

As happened in the pressure effect analysis, the correlation between $\langle \alpha^{MSS} \rangle$ and $\langle R_C \rangle$ associated with HBT directional channels is significantly improved by taking into account a dependence on the zenith angle of each channel. However, different to the pressure case, we need to consider a proportional relation between $\langle \alpha^{MSS} \rangle$ and $cos(Z)$. As shown in Figure 11-A, the set of points associated with HBT do not present a clear linear format without considering a relation with the zenith angle. The quasi-circular distribution of HBT points results in a Pearson Correlation Coefficient about 0.40. If we plot $\langle \alpha^{MSS} \rangle$ as a linear function of $\langle R_C \rangle cos(Z)$ as shown in Figure 11-B, the correlation is drastically improved. In this case, PCC increases to 0.88. However, the linear regression results (dashed blue lines) are similar when comparing $\langle \alpha^{MSS} \rangle$ as a function of $\langle R_C \rangle$ or $\langle R_C \rangle cos(Z)$. In both cases, the regression constant is close to -0.3 %/K while the regression coefficient changes from 0.029 to 0.036 %/($K \cdot GV$). Moreover, as shown in Figure 11-C, we find a good linear correlation between $\langle \alpha^{MSS} \rangle$ and $ln[\langle R_C \rangle * cos(Z)]$ for HBT. The PCC in this case is about 0.89, which is a little higher than that found in Figure 11-B. This result shows that $\langle \alpha^{MSS} \rangle$ found for HBT have a good correlation with both the product $\langle R_C \rangle cos(Z)$ and its natural logarithmic.

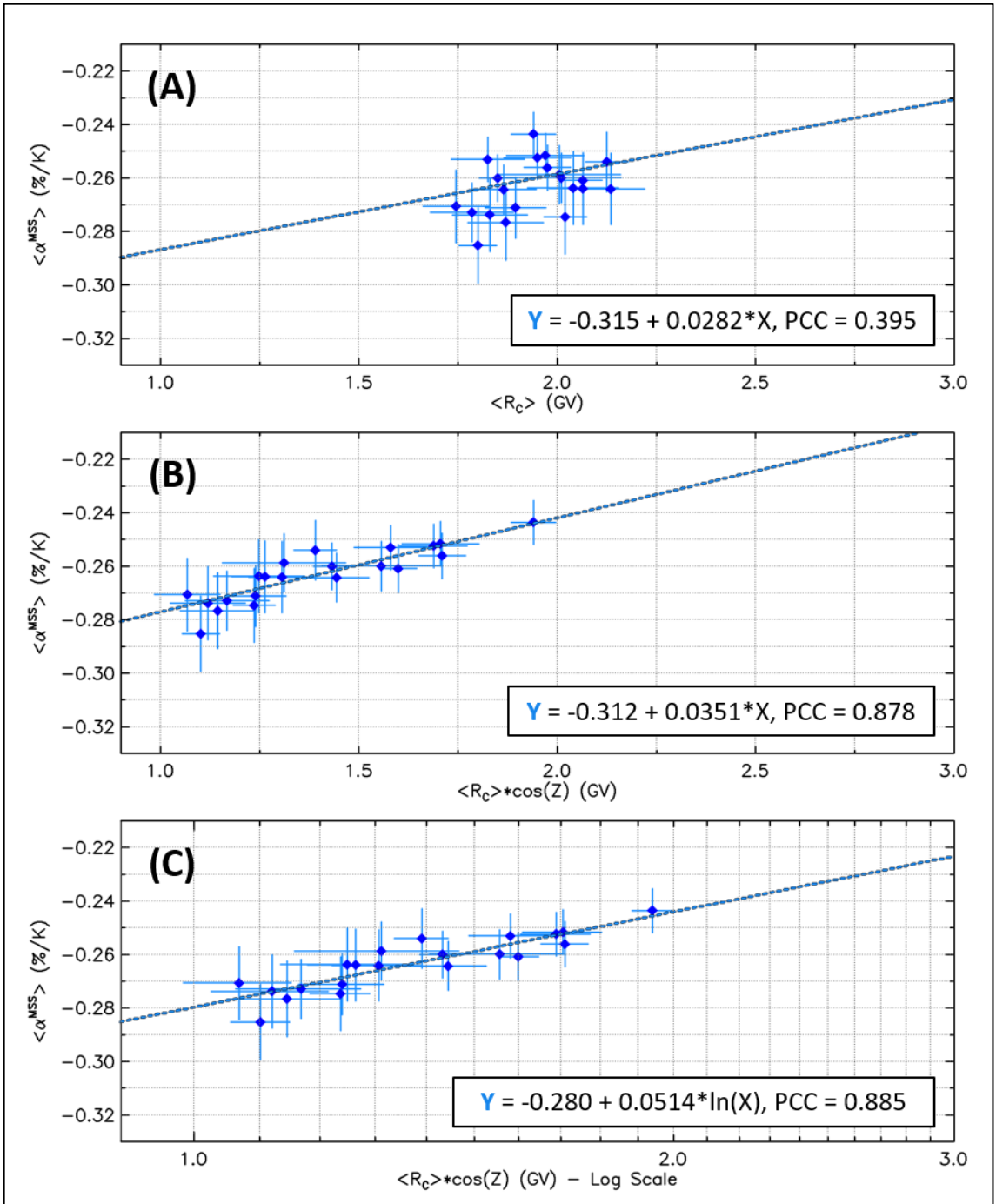

Figure 11 - Correlation of the average mass weighted temperature coefficient associated with directional channels of HBT detector with the average geomagnetic cutoff rigidity ($\langle R_C \rangle$) in panel A and with the product between $\langle R_C \rangle$ and the zenith angle ($Z$) cosine in panels B and C. The boxes in each panel show the linear regression results and the Pearson correlation coefficient (PCC) found in each case. The vertical and horizontal bars along with each point represent the uncertainties associated with $\langle \alpha^{MSS} \rangle$ and $\langle R_C \rangle$, which are not considered in the dashed curves calculation. Only $\langle \alpha^{MSS} \rangle$ points that have errors lower than 0.015%/K are shown. The values used to produce this figure are available in the Supporting Information tables.

Figure 12 shows the results when analyzing the temperature coefficients relationship with cutoff rigidity and zenith angle for all GMDN data. At first glance, we do not see any significant changes in the slope of the regression lines obtained for each detector. The lines found for SMS (green), NGY (red) and KWT (brown) present similar slopes while HBT (blue) presents a slope about ten times larger. The correlation coefficients, on the other hand, are changed from the case in Figure 10 in a complex way. For KWT and HBT, we find an improvement of PCC while we find PCC decreased for SMS and NGY. Particularly for NGY, the PCC is decreased from 0.75 to 0.66, although we cannot see significant difference between linear alignments of NGY points (red triangles) shown in Figures 10 and 12. On the other hand, we see by eye a very clear difference between HBT points (blue diamonds) alignments. As already discussed, the set of points of this detector only shown a very clear linear distribution after considering that $\langle \alpha^{MSS} \rangle$ is related to $\langle R_C \rangle cos(Z)$. Moreover, when considering this relation, we found a higher correlation coefficient value for all detectors together. The average PCC for all four detectors in this case is about 0.781 while it is about 0.697 when considering a relation only with $\langle R_C \rangle$. In this way, we consider that, in general, the mass weighted temperature coefficient presents a better relation to the product between the effective geomagnetic cutoff rigidity and the cosine of the zenith angle. Finally, it is also important to notice that, in this case, HBT and SMS data can be linked. As gray dashed curve on Figure 12 shows, the possible natural logarithmic relation found for HBT data seems to align with the SMS points set linear distribution. Thus, we can find a natural logarithmic relation (black curve) that fits very well HBT and SMS data and links both individual linear relations (blue and green dashed lines). We found a Pearson Correlation Coefficient of 0.982 when assuming a linear relation between $\langle \alpha^{MSS} \rangle$ and $ln[\langle R_C \rangle cos(Z)]$ obtained for these two detectors.

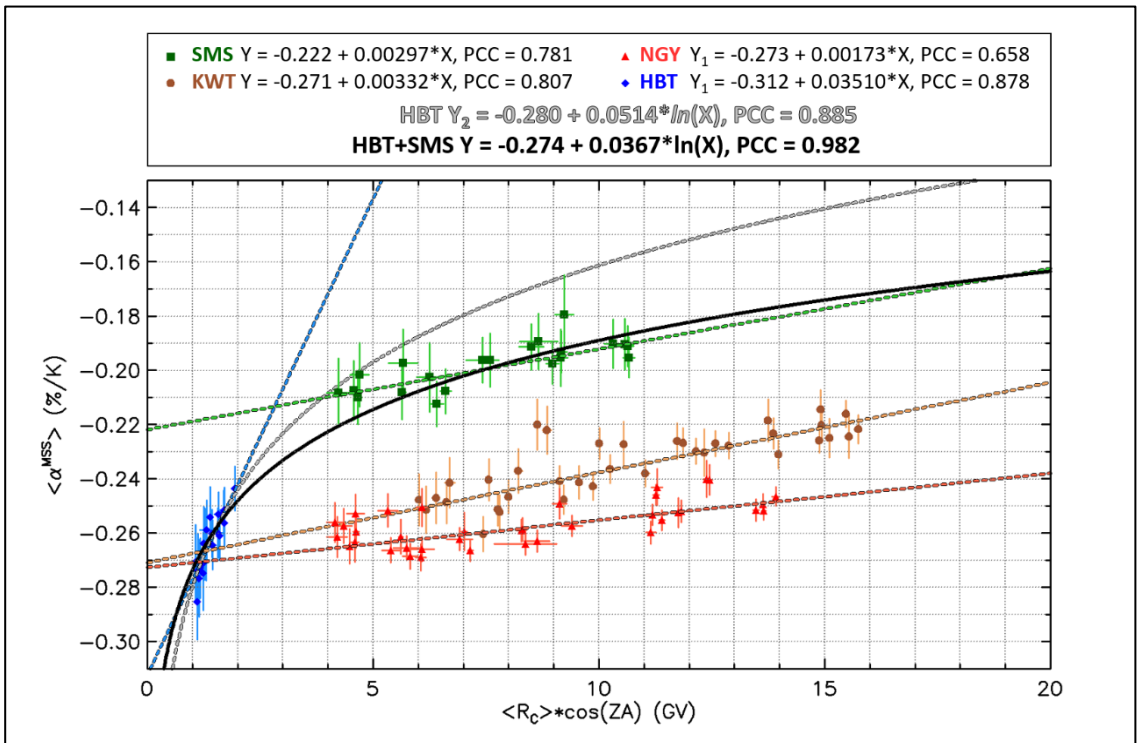

Figure 12 - Correlation between the average mass weighted temperature coefficient ($\langle \alpha^{MSS} \rangle$) and the product between the average geomagnetic cutoff rigidity ($\langle R_C \rangle$) and the zenith angle ($Z$) cosine associated with GMDN directional channels. The top boxes show the linear regression results and the Pearson correlation coefficient (PCC) found using data from each detector individually. The vertical and horizontal bars along with each point represent the uncertainties associated with $\langle \alpha^{MSS} \rangle$ and $\langle R_C \rangle$, which are not taken into account in the dashed curves calculation. Only $\langle \alpha^{MSS} \rangle$ points that have errors lower than 0.015%/K are shown. The values used to produce this figure are available in the Supporting Information tables.

Considering that (I) the values of $\langle \alpha^{MSS} \rangle$ for HBT and SMS can be expressed by a single linear function of $ln[\langle R_C \rangle cos(Z)]$ and (II) the linear functions for SMS, NGY and KWT detectors have similar slopes, we can assume that set of points of the last two are shifted below by some unknown effect. It therefore seems that there is an effect changing all values of $\langle \alpha^{MSS} \rangle$ obtained for NGY and KWT in a way that their correlation with $\langle R_C \rangle cos(Z)$ results a smaller linear coefficient than that obtained for SMS.

When correcting this effect, if we add about 0.05 %/K to all $\langle \alpha^{MSS} \rangle$ determined for KWT and NGY, we would have values similar to those found for SMS. Doing this allows, for all four detectors, $\langle \alpha^{MSS} \rangle$ to be expressed by a single linear function of $ln[\langle R_C \rangle * cos(Z)]$. In this case, $\langle \alpha^{MSS} \rangle$ found for KWT and NGY detector shown in Figure 12 would change from around -0.24 and -0.26 %/K to -0.19 and -0.21 %/K, respectively, and fit to the continuous black curve. The main question is what local aspect of the detectors can be related with these changes.

As we can see in Figure 1, HBT and SMS detectors, which are not influenced by our hypothetical effect, are in the Southern Hemisphere while KWT and NGY (both affected) are in the Northern Hemisphere. We thus assume an effect linked to the geographical latitude of detector site and consider a hypothetical mass weighted temperature coefficient ($\delta^{MSS}$) for each GMDN directional channel, given by the following:

$$\delta^{MSS} = \langle \alpha^{MSS} \rangle + K * sin(Lat_D) \qquad (6)$$

where $\langle \alpha^{MSS} \rangle$ is the average mass weighted temperature coefficient found for each GMDN detector, i.e., those shown in Figure 9. $K$ is an arbitrary positive constant and $Lat_D$ is the geographical latitude of the detector site. Through this equation, the temperature coefficients corrected by our hypothetical "latitude effect" will be higher than those experimentally found for KWT and NGY. The opposite occurs for SMS because of their negative value of $Lat_D$. In this way, the corrected values of temperature coefficients ($\delta^{MSS}$) found for these detectors tends to be similar depending on the value of $K$. In order to find this constant value, we consider:

$$2\langle \delta^{MSS} \rangle_{SMS} - \langle \delta^{MSS} \rangle_{NGY} - \langle \delta^{MSS} \rangle_{KWT} = 0 \qquad (7)$$

where $\langle \delta^{MSS} \rangle_{SMS}$, $\langle \delta^{MSS} \rangle_{NGY}$ and $\langle \delta^{MSS} \rangle_{KWT}$ are the mean hypothetical temperature coefficient found for SMS, NGY and KWT detector, respectively. These average values are obtained as follows:

$$\langle \delta^{MSS} \rangle_D = \frac{1}{N_{D:c}} \sum_c \langle \alpha^{MSS} \rangle_{D:c} + K * sin(Lat_D) \qquad (8)$$

where $\langle \delta^{MSS} \rangle_D$ is the mean hypothetical temperature coefficient found for the detector "$D$"; $N_{D:c}$ is the total number of directional channels of this detector; $\langle \alpha^{MSS} \rangle_{D:c}$ is the average mass weighted temperature coefficient found for the directional channel "$c$" of this detector; $K$ is the arbitrary positive constant that we want to obtain; and $Lat_D$ is the latitude of detector "$D$" site. In a few words, $\langle \delta^{MSS} \rangle_D$ is equivalent to the average value of temperature coefficients found for a detector plus a term based on the latitude effect. The first right term of Eq. (8) can be calculated through the data shown in Figure 9. Thus, using the corresponding $Lat_D$ (-29.44° for SMS, +35.15° for NGY and +29.37° for KWT), we found that $\langle \delta^{MSS} \rangle_{SMS} = -0.1981 - 0.4915 * K$; $\langle \delta^{MSS} \rangle_{NGY} = -0.2547 + 0.5757 * K$; and $\langle \delta^{MSS} \rangle_{KWT} = -0.2339 + 0.4904 * K$. Using these three expressions in equation (7), we found $K$ is 0.0488.

Finally, as shown in Figure 13, we see that the points of SMS (green squares), KWT (brown circles) and NGY (red triangles) present a small scatter about the fitted line after applying the latitude-based adjustment. Moreover, when also applying this adjustment for HBT data (using the $K$ value shown above and considering that $Lat_{HBT}$ is -43.00°), we see a good alignment of all points along the black curve, which is obtained through a linear regression between $\langle \alpha^{MSS} \rangle + K * sin(Lat_D)$ and $ln[\langle R_C \rangle * cos(Z)]$. The Pearson correlation coefficient found in this case is as high as 0.952. In this way, we can conclude that our hypothetical latitude effect reproduces the observed data very well. In other words, we can say that the $\langle \alpha^{MSS} \rangle$ obtained through GMDN data analysis can be directly associated with the observation site latitude.

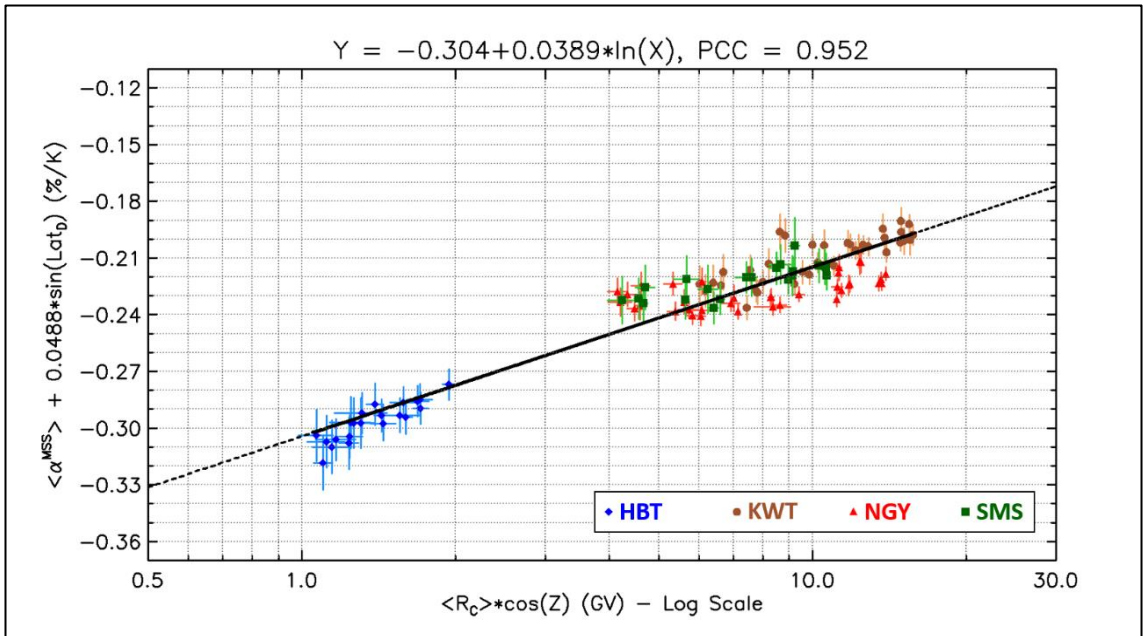

Figure 13 - Correlation between the average mass weighted temperature coefficient ($\langle \alpha^{MSS} \rangle$) added to a parameter based on latitude of the observation site ($Lat_D$) and the average geomagnetic cutoff rigidity ($\langle R_C \rangle$) multiplied by cosine of the zenith angle ($Z$) found for each GMDN directional channel. The vertical and horizontal bars along with each point represent the uncertainties associated with $\langle \alpha^{MSS} \rangle$ and $\langle R_C \rangle$, which are not considered in the black line calculation. Only $\langle \alpha^{MSS} \rangle$ points that have errors lower than 0.015%/K are shown. The values used to produce this figure are available in the Supporting Information tables.

Thus, we need to consider that the temperature effect on secondary muons has a dependence on latitude, which is hitherto unknown. Another possibility is the existence of an external influence acting together that related to the temperature effect and introducing an additional seasonal variation in the muon count rates enhancing or reducing the seasonal variation associated with the temperature effect in the both hemispheres. In this way, we would observe a dependence on latitude when analyzing the temperature effect without considering this effect. Further analysis of this and other hypotheses about the origin of the latitude effect found in this study will be undertaken in future work.

**4 - Summary**

Cosmic rays are high-energy charged particles hitting the Earth's atmosphere with a quasi-isotropic flux that is modulated by solar and interplanetary phenomena and can be used as a space weather forecast and monitoring tool. After being deflected by the geomagnetic field, they interact with atmospheric nuclei generating secondary cosmic ray particles such as muons and neutrons. Both present temporal variations related to atmospheric pressure change and, for muons, to atmospheric temperature profile alterations.

The pressure effect is observed as an anticorrelation between cosmic ray intensity and ground-level atmospheric pressure variations. This can be simply explained as a result of the increasing absorption in the atmosphere due to the increasing mass of the atmospheric above the detector. Besides this absorption effect, muon generation and decay in the atmosphere are also responsive to variations related to atmospheric pressure (Sagisaka, 1986; Dorman, 2004).

The temperature effect on muon intensity observed at ground-level is observed as a seasonal variation in an apparent anti-correlation with the temperature measured at ground. The main origin of this effect is the longer path that muons must travel before reaching the ground in the expanded atmosphere during the summer, resulting in increased likelihood of muon decay. Besides this temperature influence on muon decay, a relation with muon production is expected manly when analyzing data from muon detector with high-energy threshold (Dorman, 2004).

Many works have analyzed both atmospherics effects on muon detector data in assorted ways. Regardless of the method chosen, in general, these effects are associated with production, absorption and decay processes of secondary cosmic rays in the atmosphere. We can consider that higher energy primary particles will generate higher energy secondary particles that will be less affected by pressure and temperature effects. Thus, we can analyze their relation with energy by comparing how they change according to the primary particle's geomagnetic cutoff rigidity ($R_C$). Past studies found that the pressure effect becomes weaker as this parameter increases when analyzing neutron monitor data. As far as we know, there are no published analyses of pressure and temperature effect in relation to cutoff rigidity using data from ground-level muon detectors.

In this work, we analyzed the pressure and temperature effects on muon intensity observed at ground-level in different directions and how both are related to the value of $R_C$ associated with each field of view orientation. To do that, we used the Global Muon Detector Network (GMDN) and Global Data Assimilation System (GDAS) data from January 2007 to December 2016. The latter provides the atmospheric temperature vertical profiles that are compiled in a single value, taking into account the air mass of each measurement level using the Mass Weighted Method. The former provides ground atmospheric pressure measurements and muon intensity at different incidence directions through detectors located in Hobart (HBT), Australia, Kuwait City (KWT) Kuwait, Nagoya (NGY), Japan and Sao Martinho da Serra (SMS), Brazil. In total, HBT, SMS, NGY and KWT ground-level muon detectors observe the cosmic ray intensity in 49, 119, 121 and 529 different directions, respectively. In this work, we analyzed data from KWT by grouping 3x3 neighboring channels to decrease their count rate error. Therefore, its 529 channels reduce to 49. Moreover, for all detectors, we only used directional channels whose count rate error was less than 1% as at December 31, 2008.

We calculated the average $R_C$ for each GMDN directional channel used in this work through annual values obtained by using the Smart et al. (2000) model considering DGRF (Definitive Geomagnetic Reference Field) and IGRF-12 (International Geomagnetic Reference Field) coefficients corresponding to the period analyzed in this work. Due to the location of the HBT detector, the values found for it are about 10 GV lower than those obtained for other detectors. Moreover, it is possible to notice a clear east-west asymmetry in the geomagnetic cutoff rigidity found for all GMDN detectors that is associated with the well-known East-West geomagnetic effect on cosmic rays.

Using ground-level atmospheric pressure variation ($\Delta P$) and data from each directional channel, we obtained monthly values of barometric coefficient ($\beta$). Considering only periods where the correlation coefficient between GMDN vertical channels intensity variation and $\Delta P$ are higher than 0.7, we calculated average values of $\beta$ for each GMDN directional channel. Similar to what we observed for the cutoff rigidity analysis, we clearly noticed: (I) an east-west difference (channels looking to west are more influenced by pressure changes than those pointing to east) mainly in HBT, NGY and SMS data; and (II) that HBT detector data are much more influenced by the barometric effect when compared to other detectors. Moreover, we also found, mainly in HBT, an appreciable variation of $\beta$ according to the inclination (zenith angle) of each directional channel field of view. The higher the zenith angle, the higher the pressure coefficient magnitude, i.e., the higher the barometric effect.

Comparing average values of barometric coefficient and cutoff rigidity associated with each directional channel, we found a good natural logarithmic relation with a correlation coefficient of 0.960. Despite this good correlation found using all data, we noticed that HBT data set does not present a good correlation with the cutoff rigidity when analyzed alone. This situation is greatly improved by including a factor based on the zenith angle associated with each directional channel. After that, the correlation coefficient found using all data changes to 0.994. As a result, we can say that the barometric coefficient ($\beta$) relation with primary particle geomagnetic cutoff rigidity ($R_C$) and zenith angle ($Z$) values observed by GMDN data analysis can be summarized by the following expression:

$$\beta(R_C, Z) = -0.173 + 0.0187 * ln\left[\frac{R_C}{cos(Z)}\right] \tag{9}$$

where $\beta$ is given in %/hPa, $R_C$ in GV and $Z$ in degrees.

This expression, as well as the results found in this work, indicates that the pressure effect on secondary muons increases as primary particles geomagnetic cutoff rigidity reduces. Furthermore, for a fixed value of $R_C$, we can say that the barometric effect tends to decrease ($\beta$ tends to zero) as the zenith angle increases (as the field of view directs toward the horizon).

In order to analyze the temperature effect, we calculated the mass weighted temperature coefficients ($\alpha^{MSS}$) for each GMDN directional channel. This parameter is calculated by comparing pressure corrected muon intensity recorded in each channel with atmospheric temperature data. First, we obtained yearly values of mass weighted temperature coefficient. Later, we computed mean values using only years with a significant (> 0.7) correlation coefficient between vertical channel muon intensity corrected by pressure and $\alpha^{MSS}$ deviation. By analyzing these average values, we found a detector-dependent difference. The magnitude of coefficients found for HBT and NGY detectors are higher than those obtained using KWT data that, in turn, are higher when compared to SMS. By analyzing data associated with each detector individually, we noticed: (I) an east-west asymmetry for all detectors and (II) a variation according to the directional channel field of view zenith angle mainly in HBT data. Then, as in the pressure effect case, we found that the temperature effect tends to be weaker on channels looking to east than those looking to west. However, different from the pressure case, the temperature effect seems to be stronger for more inclined channels than for those pointing vertically.

Comparing average values of mass weighted temperature coefficient ($\alpha^{MSS}$) and primary particles geomagnetic cutoff rigidity ($R_C$) associated with each directional channel, we could not find a common relation for all GMDN together. However, we could see some correlation between $\alpha^{MSS}$ and $R_C$ when analyzing detectors individually. A good linear relation between both can be observed in SMS, KWT and NGY data. However, while their slopes are quite similar, their y-intercept values are different. In other words, the change of $\alpha^{MSS}$ as $R_C$ varies has the same slope in these three detectors, but an unknown factor adds a different y-intercept value for each detector. By analyzing HBT data, we noticed that the correlation increases significantly when including a dependency with zenith angle ($Z$), similarly to what happens in the pressure effect analysis. However, we needed to consider a proportional relation with the zenith angle cosine. Moreover, we found that $\alpha^{MSS}$ and $R_C cos(Z)$ obtained for HBT detector's directional channels can be well correlated by a natural logarithmic relation. Considering this, we see that HBT and SMS can be well described by a single correlation curve. Because these two detectors (HBT and SMS) are in the southern hemisphere and SMS agrees with northern detectors (KWT and NGY) except for y-intercept values, we assumed a local influence related to the sine of the geographic latitude of each detector's site. Analyzing SMS, KWT and NGY data, we found a proportionality constant that along with the latitude sine dependence joins the sparse data of these three detectors (reduces the differences in their y-intercept). Finally, after applying this latitude adjustment, we found a single relation for the all detectors together that presents a very good correlation coefficient (of 0.952). In this way, we can say that, based on our experimental analysis of muon intensity observed by the GMDN, the mass weighted temperature coefficient ($\alpha^{MSS}$) given in %/K can be related to the primary particle geomagnetic cutoff rigidity ($R_C$), in GV, the zenith angle ($Z$), in degrees, and the geographical latitude of the observation site ($Lat_D$), in degrees, as follows:

$$\alpha^{MSS}(R_C, Z, Lat_D) = -0.304 + 0.0389 * ln[R_C * cos(Z)] - 0.0488 * sin(Lat_D). \quad (10)$$

Thus, from the results found in this study and summarized by this expression, we can say that the temperature effect on ground muon detectors measured by the absolute value of $\alpha^{MSS}$ slowly decreases as $R_C$ increases. In addition, for a fixed value of $R_C$, we can say that the temperature effect tends to increase ($\alpha^{MSS}$ becomes more negative) as a directional channel field of view deviates from the vertical (zenith angle increase). Furthermore, we can say that the temperature effect presents an unexpected relation to the latitude of the detector site in a way that it increases when moving from South to North.

## 5 - Final Remarks

In this study, we observed correlations between atmospheric effects (pressure and temperature) and the natural logarithm of the primary particles' geomagnetic cutoff rigidity associated with each GMDN directional channel. Both correlations are improved after also considering a relation with the secant (pressure) or cosine (temperature) of zenith angle associated with each channel's field of view. Finally, in the temperature effect analysis, we only found a single relation for all GMDN detectors together when considering also a relation with the sine of the geographic latitude of each detector.

In general, on many theoretical formulations of atmospheric coefficients, only the muon energy threshold and the zenith angle associated with each observational direction are considered (Maeda, 1960; Sagisaka, 1986; Dorman, 2004; Dmitrieva et al., 2011). In our case, all GMDN detectors have the same vertical muon energy threshold of about 0.3-0.4 GeV, so we do not expect differences between atmospheric coefficients across the detectors due to this parameter. Concerning the zenith angle, it is important to note that theoretical relations of atmospheric coefficients with this parameter are much more complex than the experimental expressions found in this work. However, theory predicts that pressure coefficients tend to zero with zenith angle increase, see Figure 5.1.1-A of Dorman (2004) for instance. In the temperature case, the theoretical coefficients tend to be more negative as the zenith angle increases, see Figures 8-9 and 11-12 of Dmitrieva et al. (2011). In our experimental atmospheric coefficient analysis, we notice a similar behavior in both cases. However, we need to wait for further studies to understand how the simple relations found in this work relate to the complex zenith angle dependences predicted by existing theories.

Several latitude survey experiments show that the observed muon intensity presents a strong variation related to $R_C$ for values higher than 4 GV (see Dorman (2009) and references therein).

However, theoretically the hard-muon sea level intensity can be strongly affected by the geomagnetic field from $R_C$ about 1 GV depending on the phase factor and energy loss coefficients chosen. In addition, Allkofer et al. (1975), Kremer et al. (1999), Grieder (2001) and Cecchini & Spurio (2012) and references therein pointed out that the geomagnetic effect is very important even for low-energy muons at sea level. They found differences in the spectra, differential and integral intensity of low energy muons even when comparing data at lower cutoff rigidity (0.5-1 GV) regions.

Although there is no direct theoretical relation between the temperature effect and the primary particles' geomagnetic cutoff rigidity, Maeda (1960) theoretically analyzed how the geomagnetic deflection of secondary muons influences the temperature effect on their intensity observed at ground. Maeda (1960) predicted that the temperature effect on muon detectors with low energy threshold would be stronger in the west direction than in the east due to the differences between energy spectra of positive and negative muons at ground. In our analysis we found a similar result: the temperature coefficients associated with western directional channels are more negative (deviating from zero) than those associated with eastern channels. Considering the GMDN detector characteristics, we assume that these detectors have a low muon energy threshold. In this way, the relations of atmospheric coefficients with cutoff rigidity and zenith angle found in this work can be related to a direct geomagnetic influence on secondary muons. Thus, the primary particle particles' cutoff rigidity can be acting as a proxy of the geomagnetic deflection of secondary muons. This hypothesis will be studied in detail in future works.

It is also important to say that the relation between temperature coefficients and the latitude of the observation site found in this analysis does not appear in the theoretical studies done by Maeda, (1960), Sagisaka (1986), Dorman (2004) and Dmitrieva et al. (2011). However, Mendonça et al. (2016a) have already shown that the theory by Sagisaka (1986) overestimated temperature effect for HBT and SMS (southern hemisphere) detectors. The seasonal variation observed in the vertical directional channel of these detectors has a significantly smaller amplitude than that expected by Sagisaka (1986). On the other hand, the theoretical and observed seasonal variation in NGY and KWT (northern hemisphere) detectors are similar. Lastly, it is relevant to mention that this latitudinal relation may not be directly related to the temperature effect. It is possible that it is a consequence of an unknown or disregarded external influences acting together with the temperature effect. Thus, the origin of this relation with latitude needs to be studied in future work.

The main evidence for the atmospheric coefficients logarithmic relation with the geomagnetic cutoff rigidity and zenith angle comes from HBT data. The larger negative coefficients found for this detector, when compared to the others, evidence the logarithmic relation with cutoff rigidity. Similarly, the atmospheric coefficients variation with zenith angle are also more visible in HBT data. The logarithmic relation with the cutoff rigidity and zenith angle are also present in other GMDN detectors data analyzed in this work. However, it would be difficult to notice them if we disregard HBT data. Moreover, this logarithmic relation could be considered as counterintuitive, because muon detectors have only very small responses to low rigidity primaries around Rc such as those covered by HBT detector. Thus, it is important, for further studies, to have more data from GMDN-like muon detectors with geomagnetic cutoff rigidity range similar to HBT to better analyze this logarithmic relation.

In this work, we analyzed how the temperature effects can be related to cutoff rigidity and zenith angle associated with GMDN directional channels. In this way, we analyzed how these coefficients change in specific ranges: between 2-20 GV in the former and 0-62 degrees in the latter. Therefore, results obtained in this work cannot be applied in different ranges of cutoff rigidity and zenith angles without analyzing data measured at these ranges. Moreover, even when analyzing data observed by new detectors in the same ranges, it is necessary to take into account how the setup of these detectors is different from the GMDN ones. As discussed above, the GMDN detectors are designed to have a similar value of muon energy threshold. Since the atmospheric effects depend on this parameter, the results found in this work cannot be directly applied on muon detectors data with different muon energy threshold.

Finally, it is important to mention that the objective of this observational manuscript is to show the relation of atmospheric coefficients with geomagnetic cutoff rigidity and zenith angle found using the GMDN data. Due to the complexity of the atmospheric effects on secondary cosmic rays, we believe that it cannot help to be very speculative to discuss possible physical mechanisms responsible for the relations observed in this work without waiting for proposed further analysis.

## 6 - Acknowledgements

This work is supported in part by the joint research programs of the Institute for Space-Earth Environmental Research (ISEE), Nagoya University and the Institute for Cosmic Ray Research (ICRR), University of Tokyo. The observations are supported by Nagoya University with the Nagoya muon detector, by INPE and UFSM with the São Martinho da Serra muon detector, and by the Australian Antarctic Division with the Hobart muon detector. Observations with the Kuwait City muon detector are supported by project SP01/09 of the Research Administration of Kuwait University. R.R.S.M. thanks the China-Brazil Joint Laboratory for Space Weather. A.D.L. thanks CNPq for grant 304209/2014-7. C.R.B. thanks grants #2017/21270-7 and #2014/24711-6, Sao Paulo Research Foundation (FAPESP). E.E. thanks CNPq (PQ 302583/2015-7) and FAPESP (2018/21657-1). Global Muon Detector Network data are available at the website (http://cosray.shinshu-u.ac.jp/crest/DB/Public/main.php) of the Cosmic Ray Experimental Team (CREST) of Shinshu University. The authors gratefully acknowledge the NOAA Air Resources Laboratory (ARL) for the provision of GDAS data used in this publication, which are available at READY website (http://www.ready.noaa.gov). The data generated in the present analysis are available in the Supporting Information and in a public domain repository through the link https://doi.org/10.6084/m9.figshare.9939554. Finally, authors thank the reviewers for their constructive comments and suggestions, which helped to improve the quality of the paper.

## 7 - References

1. Adamson, P., et al., 2010. Observation of muon intensity variations by season with the MINOS far detector. Physical Review D 81. [weblink]
2. Allkofer et al., 1975. The Low-Momentum Muon Spectrum near the Equator. Lettere Al Nuovo Cimento 12, n. 4, 107–110. [weblink]
3. Ambrosio, M., et al., 1997. Seasonal variations in the underground muon intensity as seen by MACRO. Astroparticle Physics 7, 109–124. [weblink]
4. An, F.P., et al., 2018. Seasonal variation of the underground cosmic muon flux observed at Daya Bay. Journal of Cosmology and Astroparticle Physics 2018, 001–001. [weblink]
5. Bazilevskaya, G.A., et al., 2014. Solar Cycle in the Heliosphere and Cosmic Rays. Space Science Reviews 186, 409–435. [weblink]
6. Belov, A.V., et al., 2003. Cosmic ray anisotropy before and during the passage of major solar wind disturbances. Advances in Space Research 31, 919–924. [weblink]

7. Berkova, M.D., et al., 2011. Temperature effect of the muon component and practical questions for considering it in real time. Bulletin of the Russian Academy of Sciences: Physics 75, 820–824. [weblink]
8. Braga, C.R., et al., 2013. Temperature effect correction for the cosmic ray muon data observed at the Brazilian Southern Space Observatory in São Martinho da Serra. Journal of Physics: Conference Series 409, 012138. [weblink]
9. Cane, H.V., 2000. Coronal Mass Ejections and Forbush Decreases. Space Science Reviews 93, 55–77. [weblink]
10. Cecchini, S., Spurio, M., 2012. Atmospheric Muons: Experimental Aspects." Geoscientific Instrumentation, Methods and Data Systems 1, 185-196. [weblink]
11. Dmitrieva, A.N., et al., 2011. Corrections for temperature effect for ground-based muon hodoscopes. Astroparticle Physics 34, 401–411. [weblink]
12. Dorman, L.I., 2004. Cosmic Rays in the Earth's Atmosphere and Underground, Astrophysics and Space Science Library. Springer Netherlands, Dordrecht. [weblink]
13. Dorman, L.I., 2009. Cosmic Rays in Magnetospheres of the Earth and Other Planets, Astrophysics and Space Science Library. Springer Netherlands, Dordrecht. [weblink]
14. Dorman, L.I., 2012. Cosmic rays and space weather: effects on global climate change. Annales Geophysicae 30, 9–19. [weblink]
15. Duperier, A., 1951. On the Positive Temperature Effect of the Upper Atmosphere and the Process of Meson Production. Journal of Atmospheric and Terrestrial Physics 1, 296–310. [weblink]
16. Grieder, P.K.F., 2001. Cosmic rays at earth: researcher's reference manual and data book, ed. Elsevier, Amsterdam. [weblink]
17. Herbst, K., et al., 2013. Influence of the terrestrial magnetic field geometry on the cutoff rigidity of cosmic ray particles. Annales Geophysicae 31, 1637–1643. [weblink]
18. Kremer, J., et al., 1999. Measurements of Ground-Level Muons at Two Geomagnetic Locations. Physical Review Letters 83, 4241–44. [weblink]
19. Kudela, K., 2009. On energetic particles in space. Acta Physica Slovaca. Reviews and Tutorials 59. [weblink]
20. Kudela, K., Storini, M., 2006. Possible tools for space weather issues from cosmic ray continuous records. Advances in Space Research 37, 1443–1449. [weblink]
21. Leerungnavarat, K., et al., 2003. Loss Cone Precursors to Forbush Decreases and Advance Warning of Space Weather Effects. The Astrophysical Journal 593, 587–596. [weblink]
22. Maghrabi, A., Almutairi, M., 2018. The influence of several atmospheric variables on cosmic ray muons observed by KACST detector. Advances in Space Research. [weblink]
23. Maeda, K., 1960. Directional Dependence of Atmospheric Temperature Effects on Cosmic-Ray Muons at Sea-Level. Journal of Atmospheric and Terrestrial Physics 19, 184–245. [weblink]
24. Mendonça, R.R.S. de, et al., 2013. Analysis of atmospheric pressure and temperature effects on cosmic ray measurements: Journal of Geophysical Research: Space Physics 118, 1403–1409. [weblink]
25. Mendonça, R.R.S. de, et al., 2016a. Temperature Effect in secondary cosmic rays (muons) observed at the ground: analysis of the Global Muon Detector Network Data. The Astrophysical Journal 830, 88. [weblink]
26. Mendonça, R.R.S. de, et al., 2016b. Deriving the solar activity cycle modulation on cosmic ray intensity observed by Nagoya muon detector from October 1970 until December 2012. Proceedings of the International Astronomical Union 12, 130–133. [weblink]

27. Moraal, H., 2013. Cosmic-Ray Modulation Equations. Space Science Reviews 176, 299–319. [weblink]
28. Munakata, K., et al., 2000. Precursors of geomagnetic storms observed by the muon detector network. Journal of Geophysical Research: Space Physics 105, 27457–27468. [weblink]
29. Papailiou, M., et al., 2012. Precursor Effects in Different Cases of Forbush Decreases. Solar Physics 276, 337–350. [weblink]
30. Priest, E., 2013. Magnetohydrodynamics of the Sun. Cambridge University Press, Cambridge. [weblink]
31. Rockenbach, M., et al., 2014. Global Muon Detector Network Used for Space Weather Applications. Space Science Reviews 182, 1–18. [weblink]
32. Ryan, J.M., et al., 2000. Solar Energetic Particles. Space Science Reviews 93, 35–53. [weblink]
33. Sagisaka, S., 1986. Atmospheric effects on cosmic-ray muon intensities at deep underground depths. Il Nuovo Cimento C 9, 809–828. [weblink]
34. Singh, Y.P., Badruddin, 2007. Corotating high-speed solar-wind streams and recurrent cosmic ray modulation. Journal of Geophysical Research: Space Physics 112, A05101. [weblink].
35. Smart, D.F., et al., 2000. Magnetospheric Models and Trajectory Computations. Space Science Reviews 93, 305–333. [weblink]
36. Tolkacheva, N.V., et al., 2011. Atmospheric effects in the intensity of cosmic ray muon bundles. Bulletin of the Russian Academy of Sciences: Physics 75, 377–380. [weblink]
37. Yanchukovsky, V.L., et al., 2007. Atmospheric variations in muon intensity for different zenith angles. Bulletin of the Russian Academy of Sciences: Physics 71, 1038–1040. [weblink]